\documentclass[
prd,preprint,aps,endfloats,tightenlines,
showpacs,superscriptaddress,nofootinbib,
amssymb,dvipdfmx
]{revtex4}
\usepackage{graphicx}
\usepackage{ulem}
\usepackage{color}
\begin{document}
\setlength{\baselineskip}{16pt}
%
%---------------------------------------------------------------------------
%
\title{
Scattering lengths for two pseudoscalar meson systems
}
%
%----------------------------------
\author{
Kiyoshi Sasaki
}
\affiliation{
Department of Physics, Tokyo Institute of Technology,
Tokyo 152-8551, Japan
}
%
%----------------------------------
\author{
Naruhito Ishizuka
}
\affiliation{
Graduate School of Pure and Applied Sciences, University of Tsukuba,
Tsukuba, Ibaraki 305-8571, Japan
}
\affiliation{
Center for Computational Science, University of Tsukuba,
Tsukuba, Ibaraki 305-8577, Japan
}
%
%----------------------------------
\author{
Makoto Oka
}
\affiliation{
Department of Physics, Tokyo Institute of Technology,
Tokyo 152-8551, Japan
}
%
%----------------------------------
\author{
Takeshi Yamazaki
}
\affiliation{
Kobayashi-Maskawa Institute for the Origin of Particles and the Universe, 
Nagoya University, 
Nagoya, Aichi 464-8602, Japan
}
%
%----------------------------------
%
\collaboration{PACS-CS Collaboration}
\noaffiliation
%
%----------------------------------
%
% @@ =======================================================================
%
\begin{abstract}
Scattering lengths for two pseudoscalar meson systems, 
$\pi\pi(I=2)$, $KK(I=1)$ and $\pi K(I=3/2,\ 1/2)$, 
are calculated from lattice QCD by using the finite size formula.
We perform the calculation with $N_f=2+1$ gauge configurations
generated on $32^3 \times 64$ lattice
using the Iwasaki gauge action 
and non-perturbatively ${\cal O}(a)$-improved Wilson action 
at $a^{-1} = 2.19$ GeV.
The quark masses correspond to $m_\pi = 0.17 - 0.71$ GeV.
For $\pi K(I=1/2)$ system, 
we use the variational method with the two operators, 
$\bar{s}u$ and $\pi K$, 
to separate the contamination from the higher states. 
In order to obtain the scattering length at the physical quark mass, 
we fit our results at the several quark masses 
with the formula of the ${\cal O}(p^4)$ chiral perturbation theory (ChPT) 
and that including the effects of the discretization error 
from the Wilson fermion, Wilson chiral perturbation theory (WChPT). 
We found that 
the mass dependence of our results near $m_\pi=0.17$ GeV 
are described well by WChPT but not by ChPT. 
The scattering lengths at the physical point are given as 
$a_0^{(2  )} m_\pi     =-0.04243(22)(43)$, 
$a_0^{(1  )} m_K       =-0.312  (17)(31)$, 
$a_0^{(3/2)}\mu_{\pi K}=-0.0477 (27)(20)$ and 
$a_0^{(1/2)}\mu_{\pi K}= 0.150  (16)(37)$. 
Possible systematic errors are also discussed. 
\end  {abstract}
\maketitle
%
% @@ =======================================================================
%
\section  {Introduction}
\label{sec:Introduction}
The scattering length is a key quantity
for understanding the basic properties of 
the hadron interaction at low energy.
The lattice QCD calculations of the scattering length 
for many scattering systems have been reported in the past year. 
The most of calculations employs 
the finite volume method of L\"uscher~\cite{FSM}, 
in which the scattering phase shift is related to
the energy eigenvalue on a finite volume.
In the present work, we consider to give a lattice QCD calculation 
on the scattering lengths for the 
$\pi\pi(I=2  )$, 
$ K  K (I=1  )$, 
$\pi K (I=3/2)$ and 
$\pi K (I=1/2)$ systems.

The $S$-wave $\pi\pi$ system has two isospin channels ($I=0,\ 2$). 
For the $\pi\pi(I=0)$ system, 
the time correlation function has a disconnected quark diagram. 
The statistical error of this diagram is very large and 
it makes a calculation of the scattering length very difficult 
\cite{Kuramashi.SCL,RBC.pp,Fu.pp,Fu.pp.2}. 
In the present work, we do not study this channel. 
For the $\pi\pi(I=2)$ system,
whose interaction is experimentally known to be repulsive, 
after pioneering works with the quenched approximation
\cite{Kuramashi.SCL,Sharpe.pp2,JLQCD.pp2,CPPACS1.pp2,CLQCD.pp2}, 
several authors reported the realistic calculations 
with the various formulation of the dynamical fermion 
\cite{CPPACS2.pp2,NPLQCD.pp2.0,NPLQCD.pp2,ETM.pp2,Yagi.pp2,Fu.pp,RBC.pp,Fu.pp.2}. 
For the $S$-wave $KK(I=1)$ system, 
only one calculation has been reported 
by the NPLQCD Collaboration~\cite{NPLQCD.KK1}. 
The $S$-wave $\pi K$ system has two isospin channels ($I=1/2,\, 3/2$). 
For $I=3/2$, the interaction is experimentally known to be repulsive. 
After a work with the the quenched approximation~\cite{Miao.pK32,Nagata.pK}, 
the several calculations with dynamical quarks was reported
in Ref.~\cite{NPLQCD.pK,Fu.pK,Lang.pK}.
For $I=1/2$, the interaction is known to be attractive, 
and existence of a scalar resonance with a broad width is suggested. 
The NPLQCD Collaboration evaluated the scattering length 
by using the chiral perturbation theory with the low energy constants 
obtained from the lattice calculations of the decay constants 
$f_\pi$ and $f_K$, 
and the scattering length for the $\pi K(I=3/2)$ system~\cite{NPLQCD.pK}. 
After this work, the direct calculations of the $I=1/2$ scattering length 
have been reported by some groups~\cite{Nagata.pK,Fu.pK,Lang.pK}.

Here, we note that all above calculations of the scattering length were 
performed in the quark mass range $m_\pi\ge 0.24$ GeV. 
Calculation near the physical quark mass is desired 
to evaluate the reliable results at the physical quark mass. 
In the present work, 
we calculate the scattering lengths for the 
$\pi\pi(I=2  )$, 
$ K  K (I=1  )$, 
$\pi K (I=3/2)$ and 
$\pi K (I=1/2)$ systems in $m_\pi=0.17-0.71$ GeV. 
For the $\pi K (I=1/2)$ system, 
where the existence of a scalar resonance is suggested, 
the contamination from the higher states might be non-negligible. 
To separate the contamination,
we use the variational method with the two operators, 
$\bar{s}u$ and $\pi K$. 
In order to obtain the scattering length at the physical quark mass, 
we fit our results at the several quark masses 
with the formula of the ${\cal O}(p^4)$ chiral perturbation theory (ChPT) 
and that including the effects of the discretization error 
from the Wilson fermion, Wilson chiral perturbation theory (WChPT). 
We found that the mass dependence of our results near $m_\pi=0.17$ GeV 
can be described well by WChPT, but not by ChPT.

This article is organized as follows. 
In Sec.~\ref{sec:Method_of_calculation}, 
we give 
the brief description for the L\"uscher's finite size formula and 
the calculation method of the time correlation function. 
We also give the simulation parameters. 
In Sec.~\ref{sec:Results_of_the_scattering_length}, 
we show our results of the scattering length at the several quark masses. 
In Sec.~\ref{sec:Chiral_analysis}, 
we discuss the quark mass dependence of our scattering lengths 
by using the ${\cal O}(p^4)$ ChPT and WChPT. 
In Sec.~\ref{sec:extrapolation_to_the_physical_point}, 
we evaluate the scattering lengths at the physical point 
and discuss the possible systematic errors. 
In Sec.~\ref{sec:Conclusion}, our conclusions are given. 
All calculations of the present study have been done
on the super parallel computers, 
PACS-CS and T2K-Tsukuba at the University of Tsukuba, 
and TSUBAME at the Tokyo Institute of Technology. 
The preliminary results of the present work have been reported in 
Ref.~\cite{PACSCS.SCL}. 
%
% @@ =======================================================================
%
\section  {Method of calculation}
\label{sec:Method_of_calculation}
%
% @@@ =========================================================================
%
\subsection{Scattering length}
\label{ssec:Scattering_length}
The $S$-wave scattering phase shift $\delta_0$ 
for the system of two spinless particles with mass $m_1$ and $m_2$ 
is related to the the energy eigenvalue on the finite volume by 
\begin{equation}
  \left[\, \tan\delta_0(k)/k\, \right]^{-1} = \sqrt{4\pi}\ g_{00}(k;1)
\ ,
\label{eqn:Luschers_formula}
\end  {equation}
where $k$ is the scattering momentum related to the energy 
by $E=\sqrt{m_1^2+k^2}+\sqrt{m_2^2+k^2}$. 
The function $g_{00}(k;1)$ is given by the analytic continuation of 
\begin{equation}
  g_{00}(k;z)
  = 
  \frac{\sqrt{4\pi}}{L^3}
  \sum_{{\bf p}=2\pi{\bf n}/L}({\bf p}^2 - k^2)^{-z}
  \qquad ({\bf n}\in {\mathbb Z}^3)
\ ,
\end  {equation}
which is defined for ${\rm Re}(z)>3/2$, 
where $L$ is the spatial extent. 
The scattering length is defined as, 
\begin{equation} 
  a_0 \equiv \lim_{k\to 0}\ \tan\delta_0(k)/k
\ .
\label{eqn:def_SCL}
\end  {equation}

In the case of an attractive interaction on a finite volume,
the lowest energy state has a negative $k^2$, 
so that $k$ is pure imaginary. 
In this case, for $L\to\infty$, 
two situations can be considered as 
\begin{description}
  \item[\qquad (a)]\quad $k^2\to -\kappa^2$\quad ($\kappa\in {\mathbb R}$)\ ,
  \item[\qquad (b)]\quad $k^2\to 0$\ .
\end  {description}
In (a), the system has a bound state whose binding momentum is $\kappa$. 
The $S$-matrix
\begin{equation}
  S = \mbox{e}^{2i\delta_0(k)}
    = \frac{i-\tan\delta_0(k)}{i+\tan\delta_0(k)}
\end  {equation}
has a pole at $k^2=-\kappa^2$, 
and $\tan\delta_0(k)$ converges to $-i$ in $L\to\infty$
\cite{Luscher:1985dn, Sasaki:2006jn}. 
In (b), the system has no bound state, but only scattering states. 
In general, if the system has no bound state, 
we can obtain the scattering length $a_0$
by substituting $k$ of the lowest energy state into 
Eq.(\ref{eqn:Luschers_formula}) 
and extrapolating it to $L\to\infty$, 
regardless whether the interaction is attractive or repulsive.

%
% @@@ =========================================================================
%
\subsection{Time correlation function}
\label{ssec:Time_correlation_function}

For the $\pi\pi(I=2)$ system, 
we extract the energy $E$ from the time correlation function
\begin{equation}
  G^{(2)}(t)
  =
  \langle 0|\, 
             {\Omega}^{(2)}          (t_1,t)\, 
    \overline{\Omega}^{(2)\, \dagger}(t_0  )\, 
  |0 \rangle
\ .
\label{eqn:TCor_pp2}
\end  {equation}
The operators $\Omega^{(2)}$ and $\overline{\Omega}^{(2)}$ are defined by 
\begin{eqnarray}
  \Omega^{(2)}(t_1,t) 
  &=&
  \pi^+ (t_1)\pi^+ (t) \cdot \mbox{e}^{m_\pi(t_1-t)}
\ ,
\nonumber
\\
  \overline{\Omega}^{(2)}(t_0) 
  &=&
  W_{\pi^+}(t_0+1) W_{\pi^+}(t_0)
\ ,
\label{eqn:def_Op_pp2}
\end  {eqnarray}
where $\pi^+(t)$ is the local operator 
for the $\pi^+$ meson at the time slice $t$ with the zero spatial momentum 
and $W_{\pi^+}(t)$ is the wall-source operator at the time slice $t$. 
In Eq.(\ref{eqn:def_Op_pp2}), 
the time slice of one of the wall-source operator
is shifted from the time slice of another source operator $t_0$ 
to avoid the Fierz mixing of the wall-source operators 
\cite{Kuramashi.SCL}. 
In the previous calculations, 
the time slice of the sink operators are set at $t_1=t$, 
and they simultaneously run over whole time extent. 
We call this calculation method ``method I'' in the following. 
We also employ an another method, 
where the time slice of one of the pion at $t_1$ is fixed 
and only $t$ runs over the whole time extent. 
This method is called ``method II'' in the following. 
In the method II, we need to set $t_1 \gg t$ to avoid contamination 
from higher energy states produced by the operator at $t_1$. 
For $t_0 \ll t \ll t_1$, $G^{(2)}(t)$ can be written as 
\begin{equation}
  G^{(2)}(t)
  =
  \sum_n\, 
    \langle 0   | \pi^+ | \pi \rangle\, 
    \langle \pi | \pi^+ | E_n \rangle\,  
    \langle E_n |\, \overline{\Omega}^{(2)\, \dagger}\, | 0   \rangle
    \cdot \mbox{e}^{-E_n (t-t_0)}
\ ,
\end  {equation}
where $|\pi\rangle$ is the pion state and 
where $|E_n\rangle$ is the $n$-th energy eigenstate of the two-pion state 
with the energy eigenvalue $E_n$. 
The exponential factor $\mbox{e}^{m_\pi(t_1-t)}$ 
in the definition of $\Omega^{(2)}(t_1,t)$ in Eq.(\ref{eqn:def_Op_pp2}) 
is introduced 
so that the operator $\Omega^{(2)}(t_1,t)$ 
has the same time behavior as that of the usual Heisenberg operator,
{\it i.e.},
\begin{equation}
    \langle 0 | \, \Omega^{(2)}(t_1,t  )
  = \langle 0 | \, \Omega^{(2)}(t_1,t_2)  \, {\rm e}^{ - H ( t - t_2 ) }
\qquad \mbox{ for $t_1 \gg t, t_2$ }  \ ,
\end{equation}
with the Hamiltonian $H$.

For the $KK(I=1)$ system, 
we similarly extract $E$ from the time correlation functions 
\begin{eqnarray}
  G^{(1)}(t)
  =
  \langle 0|\,  
             {\Omega}^{(1)}          (t_1,t)\, 
    \overline{\Omega}^{(1)\, \dagger}(t_0  )\, 
  |0 \rangle
\ .
\label{eqn:TCor_kk1}
\end  {eqnarray}
$\Omega^{(1)}$ and $\overline{\Omega}^{(1)}$ are defined by 
\begin{eqnarray}
  \Omega^{(1)}(t_1,t) 
  &=&
  K^+ (t_1) K^+ (t) \cdot \mbox{e}^{m_K(t_1-t)}
\ ,
\nonumber
\\
  \overline{\Omega}^{(1)}(t_0) 
  &=&
  W_{K^+}(t_0+1) W_{K^+}(t_0)
\ ,
\label{eqn:def_Op_kk1}
\end  {eqnarray}
where $K^+(t)$ is the local operator 
for the $K^+$ meson at the time slice $t$ with the zero spatial momentum 
and $W_{K^+}(t)$ is the wall-source operator at the time slice $t$.

Also for the $\pi K (I=3/2)$ system, 
we define the time correlation function 
\begin{eqnarray}
  G^{(3/2)}(t)
  =
  \langle 0|\, 
             {\Omega}^{(3/2)}          (t_1,t)\, 
    \overline{\Omega}^{(3/2)\, \dagger}(t_0  )\, 
  |0 \rangle
\ ,
\label{eqn:TCor_pk3}
\end  {eqnarray}
where 
$\Omega^{(3/2)}$ and $\overline{\Omega}^{(3/2)}$ are defined by 
\begin{eqnarray}
  \Omega^{(3/2)}(t_1,t) 
  &=&
  K^+ (t_1)\pi^+ (t) \cdot \mbox{e}^{m_K(t_1-t)}
\ ,
\nonumber
\\
  \overline{\Omega}^{(3/2)}(t_0) 
  &=&
  W_{K^+}(t_0+1) W_{\pi^+}(t_0)
\ .
\label{eqn:def_Op_pk3}
\end  {eqnarray}

Next, we construct 
the time correlation function of the $\pi K (I=1/2)$ system. 
If a resonance state exists and 
its energy is not large 
for the energy of the lowest $\pi K$ scattering state, 
the single exponential behavior for the time correlation function is seen 
only for very large time region. 
In this case, it is very difficult to extract the scattering state 
with the small statistical error. 
In order to overcome this possible problem, we use the variational method 
\cite{Diagonalization} with two kinds of operators, 
$\Omega_0^{(1/2)}$ and $\Omega_1^{(1/2)}$ 
($\overline{\Omega}_0^{(1/2)}$ and $\overline{\Omega}_1^{(1/2)}$), 
\begin{eqnarray}
           {\Omega}_0^{(1/2)}(t_1,t)
  &=&
  \frac{1}{\sqrt{3}} \left(                 K^+ (t_1  )   \pi^0 (t  )
                            - \sqrt{2}\,    K^0 (t_1  )   \pi^+ (t  )
                     \right)
                     \cdot \mbox{e}^{m_K(t_1-t)}
\ ,
\nonumber
\\
           {\Omega}_1^{(1/2)}(    t)
  &=&
  \bar{s}u  (t    )
\ ,
\nonumber
\\
  \overline{\Omega}_0^{(1/2)}(t_0)
  &=&
  \frac{1}{\sqrt{3}} \left(              W_{K^+}(t_0+1)W_{\pi^0}(t_0)
                            - \sqrt{2}\, W_{K^0}(t_0+1)W_{\pi^+}(t_0)
                     \right)
\ ,
\nonumber
\\
  \overline{\Omega}_1^{(1/2)}(t_0)
  &=&
  W_{\bar{s}u} (t_0+1)
\ ,
\label{eqn:def_Op_pk1}
\end  {eqnarray}
where 
$K^0(t)$ and $\pi^0(t)$ are the local operator 
for the $K^0$ and $\pi^0$ meson 
at the time slice $t$ with the zero spatial momentum, respectively. 
$W_{K^0}(t)$, $W_{\pi^0}(t)$ and $W_{\bar{s}u}(t)$ 
are the wall-source operators for the corresponding mesons 
at the time slice $t$, respectively. 
The exponential factor $\mbox{e}^{m_K(t_1-t)}$ 
in the definition of $\Omega_0^{(1/2)}(t_1,t)$ 
is introduced like as for the other channels.

We construct the $2\times 2$ matrix of 
a time correlation function, 
\begin{equation}
  G^{(1/2)}_{ij}(t)
  =
  \langle 0|\,         {\Omega}_i^{(1/2)}           (t_1,t)\, 
              \overline{\Omega}_j^{(1/2)\, \dagger} (t_0  )\, 
  |0 \rangle
  \qquad (\, i,\, j = 0, 1\, )\ .
\label{eqn:TCor_pk1}
\end  {equation}
In the method I 
the sink operators are set to the equal time slice, $t_1=t$, 
and they simultaneously run over whole time extent. 
For the $\pi K (I=1/2)$ system, 
we need to repeat solving quark propagators for the whole time extent 
as explained later 
and the computational costs become huge. 
Thus, we only employ the method II for this channel. 
For $t_0 \ll t \ll t_1$, 
$G^{(1/2)}_{ij}(t)$ can be written by
\begin{equation}
  G^{(1/2)}_{ij}(t)
  =
  \sum_n\ w_{in} v_{nj} \cdot \mbox{e}^{-E_n (t-t_0)}
\ ,
\end  {equation}
where
\begin{eqnarray}
  w_{0n}
  &=&
  \frac{1}{\sqrt{3}}
  \left[\, 
    \langle 0   | K^+   | K   \rangle\, 
    \langle K   | \pi^0 | E_n \rangle\, 
    - \sqrt{2}\
    \langle 0   | K^0   | K   \rangle\, 
    \langle K   | \pi^+ | E_n \rangle\, 
  \right]
\ ,
\nonumber
\\
  w_{1n}
  &=&
    \langle 0   |\,          {\Omega}_1^{(1/2)          }\, | E_n \rangle
\ ,
\nonumber
\\
  v_{nj}
  &=&
    \langle E_n |\, \overline{\Omega}_j^{(1/2)\, \dagger}\, | 0   \rangle
\ .
\end  {eqnarray}
$|K\rangle$ is the kaon state and 
$|E_n\rangle$ is the $n$-th energy eigenstate of the $\pi K$ state 
with the energy eigenvalue $E_n$. 
We assume that the lowest two states dominate 
the time correlation function in a large time region. 
With this assumption, 
we can extract the energy $E_n$ by a single exponential fit 
for two eigenvalues $\bar{G}_n(t)\ (n=0,1)$ as, 
\begin{equation}
  \bar{G}_n(t)
  = 
  \mbox{Ev}\left[\, 
    \left[\, G^{(1/2)}(t_R)\, \right]^{-1}\cdot G^{(1/2)}(t)\, 
  \right]_n
  =
  {\rm e}^{-E_n (t-t_R)}\ ,
\label{eq:barG}
\end  {equation}
where $\mbox{Ev}[M]_n$ means the $n$-th eigenvalue of the matrix $M$ 
and $t_R$ is some reference time. 
The function $\bar{G}_n(t)$ is the time correlation function 
of an optimal operator $\phi_n$ 
for the $n$-th energy eigenstate $|E_n\rangle$, 
whose property is $\langle 0|\phi_n(t)|E_m\rangle
=\delta_{nm} {\mathrm e}^{-E_n t}$.

Next, we explain 
the construction of the time correlation functions by quark propagators. 
The time correlation functions of the $\pi\pi (I=2)$ and $KK(I=1)$ systems 
are given by 
\begin{eqnarray}
  G^{(2)}(t)
  &=&
    G^{\pi\pi \to \pi\pi}_{\rm direct\, 1}(t)
  + G^{\pi\pi \to \pi\pi}_{\rm direct\, 2}(t)
  - G^{\pi\pi \to \pi\pi}_{\rm cross\,  1}(t)
  - G^{\pi\pi \to \pi\pi}_{\rm cross\,  2}(t)\ 
\ ,
\label{eqn:4pt_func_pp2}
\\
  G^{(1)}(t)
  &=&
    G^{KK \to KK}_{\rm direct\, 1}(t)
  + G^{KK \to KK}_{\rm direct\, 2}(t)
  - G^{KK \to KK}_{\rm cross\,  1}(t)
  - G^{KK \to KK}_{\rm cross\,  2}(t)\ 
\ ,
\label{eqn:4pt_func_kk1}
\end  {eqnarray}
where 
\begin{eqnarray}
  G^{\pi\pi \to \pi\pi}_{\rm direct\, 1}(t)
  &=&
  \left<\ 
  X^{\pi}( t_1 | t_0+1 )\, 
  X^{\pi}( t   | t_0   )
  \cdot
  \mbox{e}^{m_\pi\cdot ( t_1 - t )}\ 
  \right>
\ ,
\nonumber
\\
  G^{\pi\pi \to \pi\pi}_{\rm direct\, 2}(t)
  &=&
  \left<\ 
  X^{\pi}( t   | t_0+1 )\, 
  X^{\pi}( t_1 | t_0   )
  \cdot
  \mbox{e}^{m_\pi\cdot ( t_1 - t )}\ 
  \right>
\ ,
\nonumber
\\
  G^{\pi\pi \to \pi\pi}_{\rm cross\, 1}(t)
  &=&
  \left<\ 
  \mbox{Tr}
  \left[\, 
  U^{ll          }( t_1 | t_0 )\, 
  U^{ll\, \dagger}( t   | t_0 )\, 
  \right]
  \cdot
  \mbox{e}^{m_\pi\cdot ( t_1 - t )}\ 
  \right>
\ ,
\nonumber
\\
  G^{\pi\pi \to \pi\pi}_{\rm cross\, 2}(t)
  &=&
  \left<\ 
  \mbox{Tr}
  \left[\, 
  U^{ll          }( t   | t_0 )\, 
  U^{ll\, \dagger}( t_1 | t_0 )\, 
  \right]
  \cdot
  \mbox{e}^{m_\pi\cdot ( t_1 - t )}\ 
  \right>
\ ,
\nonumber
\\
  G^{KK \to KK}_{\rm direct\, 1}(t)
  &=&
  \left<\ 
  X^{K}( t_1 | t_0+1 )\, 
  X^{K}( t   | t_0   )
  \cdot
  \mbox{e}^{m_K\cdot ( t_1 - t )}\ 
  \right>
\ ,
\nonumber
\\
  G^{KK \to KK}_{\rm direct\, 2}(t)
  &=&
  \left<\ 
  X^{K}( t   | t_0+1 )\, 
  X^{K}( t_1 | t_0   )
  \cdot
  \mbox{e}^{m_K\cdot ( t_1 - t )}\ 
  \right>
\ ,
\nonumber
\\
  G^{KK \to KK}_{\rm cross\, 1}(t)
  &=&
  \left<\ 
  \mbox{Tr}
  \left[\, 
  U^{sl          }( t_1 | t_0 )\, 
  U^{ls\, \dagger}( t   | t_0 )\, 
  \right]
  \cdot
  \mbox{e}^{m_K\cdot ( t_1 - t )}\ 
  \right>
\ ,
\nonumber
\\
  G^{KK \to KK}_{\rm cross\, 2}(t)
  &=&
  \left<\ 
  \mbox{Tr}
  \left[\, 
  U^{sl          }( t   | t_0 )\, 
  U^{ls\, \dagger}( t_1 | t_0 )\, 
  \right]
  \cdot
  \mbox{e}^{m_K\cdot ( t_1 - t )}\ 
  \right>
\ .
\label{eq:eq_G_contruct}
\end  {eqnarray}
In Eq.(\ref{eq:eq_G_contruct}) 
the angle bracket refers to 
the expectation value over the gauge configurations, 
and the trace is taken for the color and spinor indices. 
The exponential factors 
$\mbox{e}^{m_\pi(t_1-t)}$ and 
$\mbox{e}^{m_K  (t_1-t)}$ come from 
the definitions of 
$\Omega^{(2)}$ in Eq.~(\ref{eqn:def_Op_pp2}) and 
$\Omega^{(1)}$ in Eq.~(\ref{eqn:def_Op_kk1}), respectively. 
The indices $l$ and $s$ 
means the up/down and strange quark, respectively. 
$X^\pi (t|t_s)$, $X^K (t|t_s)$ and $U^{f_1 f_2}(t|t_s)$ are defined by 
\begin{eqnarray}
  X^\pi ( t | t_s )
  &=&
  \sum_{\bf x}\, 
  \mbox{Tr}
  \left[\, 
    Q^{l\, \dagger}( {\bf x}, t | t_s )\, 
    Q^{l\,        }( {\bf x}, t | t_s )\, 
  \right]
\label{eqn:2pt_func_pi}
\ ,
\\
  X^K ( t | t_s )
  &=&
  \sum_{\bf x}\, 
  \mbox{Tr}
  \left[\, 
    Q^{s\, \dagger}( {\bf x}, t | t_s )\, 
    Q^{l\,        }( {\bf x}, t | t_s )\, 
  \right]
\label{eqn:2pt_func_kn}
\ ,
\\
  U_{AB}^{f_1 f_2}( t | t_s )
  &=&
  \sum_{\bf x} \sum_C\,
  Q_{CA}^{f_1\, *}( {\bf x}, t | t_s+1 )\,
  Q_{CB}^{f_2    }( {\bf x}, t | t_s   )\,
  \qquad (\, f_1, f_2 = l, s\, )
\ ,
\label{eqn:def_Uprop}
\end  {eqnarray}
with the quark propagator with the wall source 
\begin{equation}
  Q_{AB}^f ( {\bf x}, t | t_s )
  =
  \sum_{\bf y}\ 
  (D^{-1})_{AB}^f ( {\bf x}, t ; {\bf y}, t_s )
  \qquad (\, f = l, s\, )
\ ,
\label{eqn:def_Qprop}
\end  {equation}
where $A$, $B$ and $C$ refer to color and spinor indices.

\begin{figure}[htbp]
\begin{center}
\includegraphics[width=160mm]{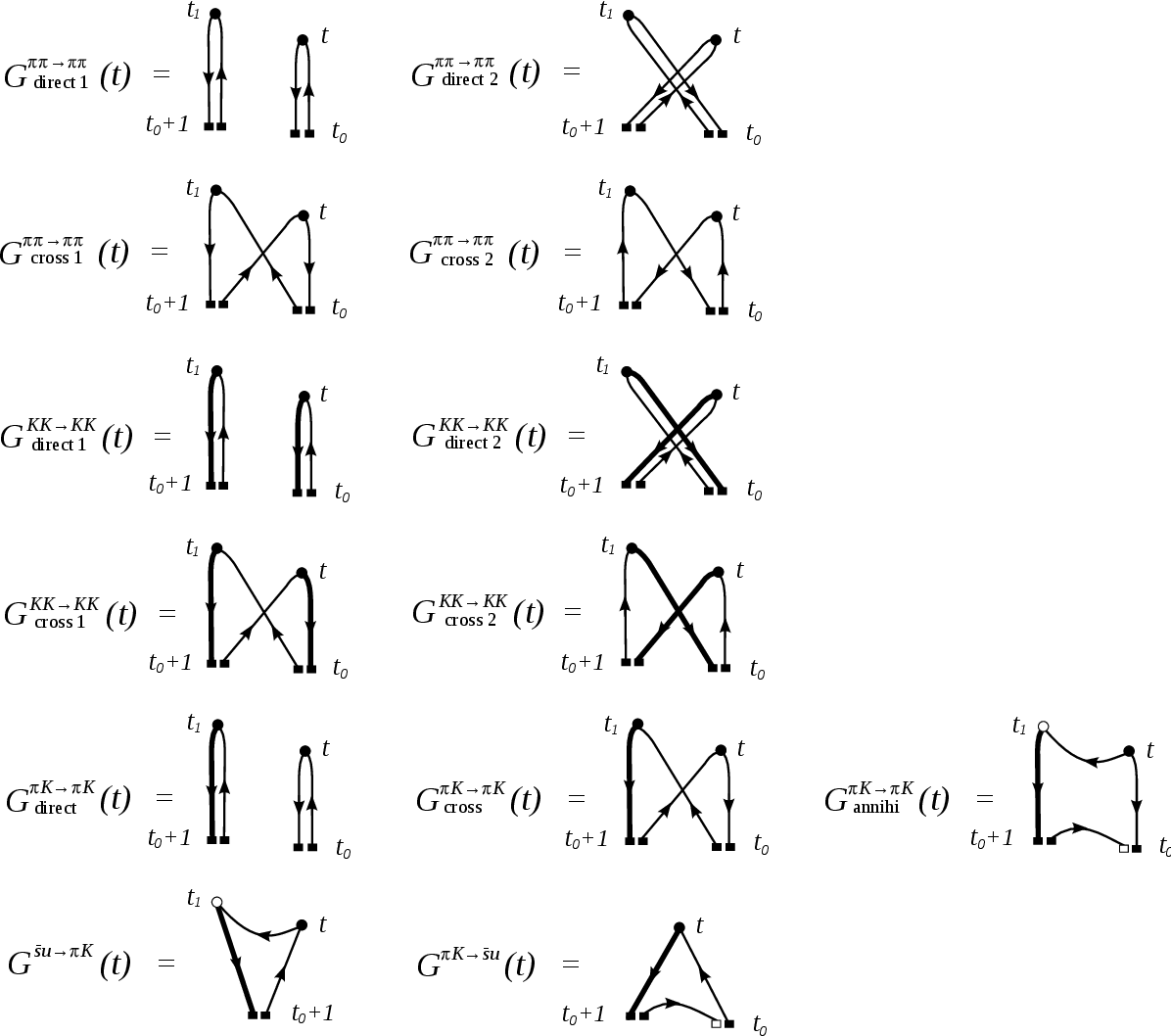}
\caption{List of the diagrams 
         employed to calculate the time correlation function
         for the $S$-wave 
         $\pi\pi(I=2  )$, 
         $ K  K (I=1  )$, 
         $\pi K (I=3/2)$ and 
         $\pi K (I=1/2)$ systems. 
         The thin (thick) lines represent 
         the up/down (strange) quark propagators.
         The time runs upward. 
         The circles are 
         the local operators for the $\pi$, $K$ and $\bar{s}u$ mesons 
         at the time slice specified in each diagram, 
         and the squares are the wall-source operators  
         for these mesons. 
         The open symbols mean the operators summed over ${\bf y}$ in 
         Eqs.(\ref{eqn:def_Vprop}) and 
             (\ref{eqn:def_Wprop}). 
         }
\label{fig:diagram}
\end  {center}
\end  {figure}

The quark diagrams for the components in Eq.(\ref{eq:eq_G_contruct})
are shown in Fig.~\ref{fig:diagram}. 
The thin (thick) lines represent
the up/down (strange) quark propagators. 
The time runs upward in the diagrams. 
The circles are 
the local operators 
for the $\pi$, $K$ and $\bar{s}u$ mesons
at the time slice specified in each diagram, 
and the squares are the wall-source operators
for these mesons.
%
%---------------------------------
%---------------------------------
%

$G^{(3/2)}(t)$ and $G^{(1/2)}_{ij}(t)$ are constructed by 
\begin{eqnarray}
  G^{(3/2)}(t)
  &=&
    G^{\pi K \to \pi K}_{\rm direct}(t)
  - G^{\pi K \to \pi K}_{\rm cross }(t)
\ ,
\\
  G^{(1/2)}_{00}(t)
  &=&
                 G^{\pi K \to \pi K}_{\rm direct}(t)
  + \frac{1}{2}\ G^{\pi K \to \pi K}_{\rm cross }(t)
  - \frac{3}{2}\ G^{\pi K \to \pi K}_{\rm annihi}(t)
\ ,
\\
  G^{(1/2)}_{01}(t)
  &=&
  - \sqrt{\frac{3}{2}}\, G^{\bar{s}u \to \pi K}(t)
\ ,
\\
  G^{(1/2)}_{10}(t)
  &=&
  - \sqrt{\frac{3}{2}}\, G^{\pi K \to \bar{s}u}(t)
\ ,
\\
  G^{(1/2)}_{11}(t)
  &=&
  G^{\bar{s}u}( t | t_0+1 )
\ ,
\end  {eqnarray}
where 
\begin{eqnarray}
  G^{\pi K \to \pi K}_{\rm direct}(t)
  &=&
  \left<\ 
  X^{K  }( t_1 | t_0+1 )\, 
  X^{\pi}( t   | t_0   )
  \cdot
  \mbox{e}^{m_K\cdot ( t_1 - t )}\ 
  \right>
\ ,
\nonumber
\\
  G^{\pi K \to \pi K}_{\rm cross}(t)
  &=&
  \left<\ 
  \mbox{Tr}
  \left[\, 
  U^{sl          }( t_1 | t_0 )\, 
  U^{ll\, \dagger}( t   | t_0 )\, 
  \right]
  \cdot
  \mbox{e}^{m_K\cdot ( t_1 - t )}\ 
  \right>
\ ,
\nonumber
\\
  G^{\pi K \to \pi K}_{\rm annihi}(t)
  &=&
  \left<\ 
  \sum_{\bf x}\, 
  \mbox{Tr}
  \left[\, 
  W^\dagger( {\bf x}, t | t_1 | t_0+1 )\, 
  V        ( {\bf x}, t | t_0+1 )\, 
  \right]
  \cdot
  \mbox{e}^{m_K\cdot ( t_1 - t )}\ 
  \right>
\ ,
\nonumber
\\
  G^{\bar{s}u \to \pi K}(t)
  &=&
  \left<\ 
  \sum_{\bf x}\, 
  \mbox{Tr}
  \left[\, 
  W^\dagger( {\bf x}, t | t_1 | t_0+1 )\, 
  \gamma_5\, 
  Q^{\, l} ( {\bf x}, t | t_0+1 )\, 
  \right]
  \cdot
  \mbox{e}^{m_K\cdot ( t_1 - t )}\ 
  \right>\ ,
\nonumber
\\
  G^{\pi K \to \bar{s}u}(t)
  &=&
  \left<\ 
  \sum_{\bf x}\, 
  \mbox{Tr}
  \left[\, 
  Q^{s\, \dagger}( {\bf x}, t | t_0+1 )\, 
  \gamma_5\, 
  V              ( {\bf x}, t | t_0+1 )\, 
  \right]\ 
  \right>
\ ,
\nonumber
\\
  G^{\bar{s}u} ( t | t_s )
  &=&
  \left<\ 
  \sum_{\bf x}\, 
  \mbox{Tr}
  \left[\, 
    \gamma_5\, 
    Q^{s\, \dagger}( {\bf x}, t | t_s )\, 
    \gamma_5\, 
    Q^{l\,        }( {\bf x}, t | t_s )\, 
  \right]\ 
  \right>
\ .
\label{eq:eq_G_contruct2}
\end  {eqnarray}
In Eq.(\ref{eq:eq_G_contruct2}) 
the exponential factor $\mbox{e}^{m_K (t_1-t)}$ comes from 
the definitions of 
$\Omega^{(3/2)}  $ in Eq.~(\ref{eqn:def_Op_pk3}) and 
$\Omega^{(1/2)}_0$ in Eq.~(\ref{eqn:def_Op_pk1}). 
$V({\bf x},t|t_s)$ and $W({\bf x},t|t_a|t_s)$ are defined by 
\begin{eqnarray}
  V_{AB}( {\bf x}, t | t_s )
  &=&
  \sum_C\ 
  Q_{AC}^l ( {\bf x}, t | t_s-1 )\, 
  \left\{\, 
  \gamma_5\, 
  \sum_{\bf y}
  Q^l ( {\bf y}, t_s-1 | t_s )\, 
  \right\}_{CB}
\ , 
\label{eqn:def_Vprop}
\\
  W_{AB}( {\bf x}, t | t_a | t_s )
  &=&
  \sum_{\bf y} \sum_C\, 
  (D^{-1})_{AC}^l ( {\bf x}, t ; {\bf y}, t_a )\, 
  \Bigl[\, 
  \gamma_5\, 
  Q^s ( {\bf y}, t_a | t_s )\, 
  \Bigr]_{CB}
\ ,
\label{eqn:def_Wprop}
\end  {eqnarray}
where the square bracket in Eq.(\ref{eqn:def_Wprop}) is taken as 
the source in solving the propagator. 
The quark diagram for the components in Eq.(\ref{eq:eq_G_contruct2})
are plotted in Fig.~\ref{fig:diagram}. 
The open symbols mean the operators summed over ${\bf y}$ in 
Eqs.(\ref{eqn:def_Vprop}) and (\ref{eqn:def_Wprop}). 
In the method I, 
we must solve $W( {\bf x}, t | t | t_0+1 )$
for each $t$ in the calculation of $G^{\pi K \to \pi K}_{\rm annihi}(t)$ 
and $G^{\bar{s}u \to \pi K}(t)$. 
On the other hand, 
in the method II, we need to solve it only once at $t_1$. 
Therefore, the computational cost is reduced 
for the $\pi K(I=1/2)$ system.

We solve four kinds of $Q$-type propagators in Eq.~(\ref{eqn:def_Qprop})
with $(f,t_s)=(l,t_0+1)$, $(s,t_0+1)$, $(l,t_0)$, $(s,t_0)$. 
We also solve one $W$-type propagator in Eq.~(\ref{eqn:def_Wprop}) 
with $(t_a,t_s)=(t_1,t_0+1)$. 
Thus, we solve $5$ quark propagators for each configuration.

% @@@ =========================================================================
%
\subsection{Simulation parameters }
\label{ssec:Simulation_parameters}
The calculations are carried out 
with $N_f=2+1$ full QCD configurations generated by the PACS-CS Collaboration
\cite{Aoki:2008sm}
using the Iwasaki gauge action at $\beta=1.90$ 
and non-perturbatively ${\cal O}(a)$-improved Wilson quark action 
with $C_{SW}=1.715$ on $32^3 \times 64$ lattice. 
The lattice cutoff is $a^{-1}=2.194(10)$ GeV ($a=0.08995(40)$ fm) 
determined from the $\Omega$-baryon mass. 
The spatial extent of the lattice is $La=2.878(13)$ fm~\cite{Aoki:2009ix}. 
The statistical error of $a$ is not included in the following analysis. 
The quark mass parameters, 
the corresponding hadron masses and 
the number of configurations 
are listed in Table \ref{tbl:mass-param}.

The quark propagators 
are calculated with the same action as the configuration generation. 
They are solved on the configurations at every 
20 trajectories for $\kappa_{ud}=0.13781$, and 
10 trajectories for the others. 
The Dirichlet boundary condition is imposed for the temporal direction and 
the periodic boundary condition for the spatial directions. 
The wall source is used 
with the gauge configurations fixed to the Coulomb gauge. 
The time slices of the source operators are $t_0=12$ and $t_0+1=13$, 
and the time slice of the fixed sink operator is set at $t_1=53$. 
We adopt $t_R=18$ as the reference time 
for the diagonalization for the $\pi K (I=1/2)$ system. 
The statistical errors are evaluated 
by the jackknife analysis with a binsize of 
110 MD time for $\kappa_{ud}=0.13781$, and 
125 MD time for the others. 
Here, the MD time is 
the number of trajectories multiplied by the trajectory length $\tau$, 
which takes 
$\tau=0.25$ for $\kappa_{ud}=0.13781$ and $0.13770$,
and $\tau=0.5 $ for other $\kappa_{ud}$.

We calculate the time correlation functions 
on the gauge configurations 
shifted by $T_\mathrm{shift}$ in the temporal direction 
and take an average of them to improve the statistics. 
We use $T_\mathrm{shift}$ listed in Table \ref{tbl:mass-param}, 
but do not include $T_\mathrm{shift}=0$ 
for the analysis of the $KK(I=1)$ system in all the quark masses. 
%
%---------------------------------
\begin{table}[htbp]
\begin{center}
\begin{tabular}{ccccccccccc}
\hline
\hline
$\kappa_{ud}$     & $\quad$ &  
$\kappa_s$        & $\quad$ &
$m_\pi$ [GeV]     & $\quad$ & 
$m_K$ [GeV]       & $\quad$ &
$N_\mathrm{conf}$ & $\quad$ &
$T_\mathrm{shift}$ \\
\hline
$0.13781$         &         & 
$0.13640$         &         &
$0.1661(58)$      &         & 
$0.5594(23)$      &         &
$154$             &         &
$0, 16, 32, 48$    \\
$0.13770$         &         & 
$0.13640$         &         &
$0.2973(23)$      &         & 
$0.5975(17)$      &         &
$800$             &         &
$0, 16, 32$ \\
$0.13754$         &         & 
$0.13640$         &         &
$0.4144(16)$      &         & 
$0.6401(13)$      &         &
$450$             &         &
$0, 16, 32, 48$    \\
$0.13727$         &         & 
$0.13640$         &         &
$0.5746(13)$      &         & 
$0.7190(12)$      &         &
$400$             &         &
$0, 16, 32, 48$    \\
$0.13700$         &         & 
$0.13640$         &         &
$0.7069(12)$      &         & 
$0.7953(12)$      &         &
$400$             &         &
$0, 16, 32, 48$    \\
\hline
\hline
\end  {tabular}
\caption{The quark mass parameters and corresponding hadron masses.
  We do not include $T_\mathrm{shift}=0$ 
  for the analysis of the $KK(I=1)$ system.
}
\label{tbl:mass-param}
%
%  Comment by K.S.
%
%   Fit ranges are chosen from plateau in the effective mass plot:
%
%   ======================================================
%      | 0.13781 | 0.13770 | 0.13754 | 0.13727 | 0.13700 |
%      |tmin|tmax|tmin|tmax|tmin|tmax|tmin|tmax|tmin|tmax|
%   ---+----+----+----+----+----+----+----+----+----+----+
%   pi | 22 | 42 | 23 | 43 | 24 | 44 | 25 | 46 | 25 | 46 |
%   K  | 22 | 42 | 23 | 43 | 24 | 44 | 25 | 46 | 25 | 46 |
%   ======================================================
%
%---------------------------------
%
\end  {center}
\end  {table}
%
% @@ ========================================================================
%
\section  {Results of the scattering length}
\label{sec:Results_of_the_scattering_length}
%
% @@@ =======================================================================
%
\subsection{Time correlation functions and effective masses}
\label{ssec:Time_correlation_functions_and_effective_masses}

\begin{figure}[htbp]
\begin{center}
\includegraphics[width=165mm]{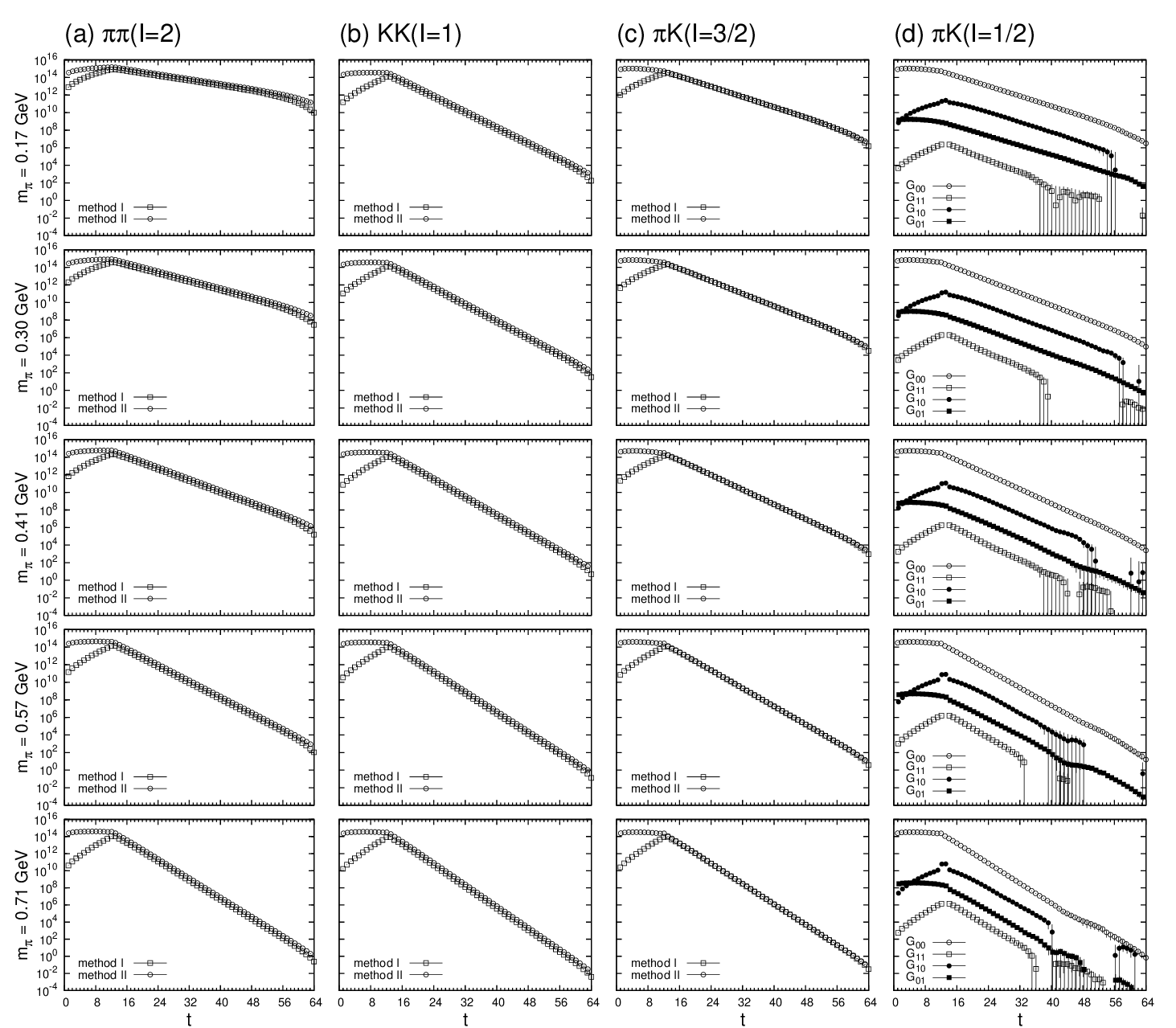}
\caption{
  The time correlation functions for 
  (a) the $\pi\pi(I=2  )$, 
  (b) the $ K  K (I=1  )$,
  (c) the $\pi K (I=3/2)$ and 
  (d) the $\pi K (I=1/2)$ systems 
  at $m_\pi=0.17$, $0.30$, $0.41$, $0.57$ and $0.71$ GeV. 
  In the columns (a), (b) and (c),
  the results of the two methods, the method I (squares) and II (circles), 
  are shown. 
  In the column (d), 
  the open symbols represent 
  the diagonal elements of $G^{(1/2)}(t)$. 
  The filled symbols represent 
  the off-diagonal elements, whose signs are reversed.
}
\label{fig:time-corr}
\end  {center}
\end  {figure}

The time correlation functions of the 
$\pi\pi(I=2  )$, 
$ K  K (I=1  )$, 
$\pi K (I=3/2)$ and 
$\pi K (I=1/2)$ systems 
which are defined 
in Eqs.(\ref{eqn:TCor_pp2}), 
       (\ref{eqn:TCor_kk1}), 
       (\ref{eqn:TCor_pk3}) and 
       (\ref{eqn:TCor_pk1}), 
are plotted in the 
columns (a), (b), (c) and (d) of Fig. \ref{fig:time-corr}, respectively.
Each row in Fig. \ref{fig:time-corr} represents 
the time correlation functions 
for $m_\pi=0.17$, $0.30$, $0.41$, $0.57$ and $0.71$ GeV. 
For the repulsive channels, 
$\pi\pi(I=2  )$, 
$ K  K (I=1  )$ and 
$\pi K (I=3/2)$, 
we employ both the method I and II 
for the calculation of the time correlation function as explained before. 
The two results are compared in the figure. 
For $\pi K (I=1/2)$ plotted in the column (d), 
the absolute values of each component of $| G^{(1/2)}(t) |$
are presented. 
As discussed in the previous section, 
we employ only the method II for this channel. 
The open symbols represent 
the diagonal elements of $G^{(1/2)}(t)$.
The filled symbols represent 
the off-diagonal elements, whose signs are reversed.

%
%---------------------------------
%---------------------------------
%

\begin{figure}[htbp]
\begin{center}
\includegraphics[height=180mm]{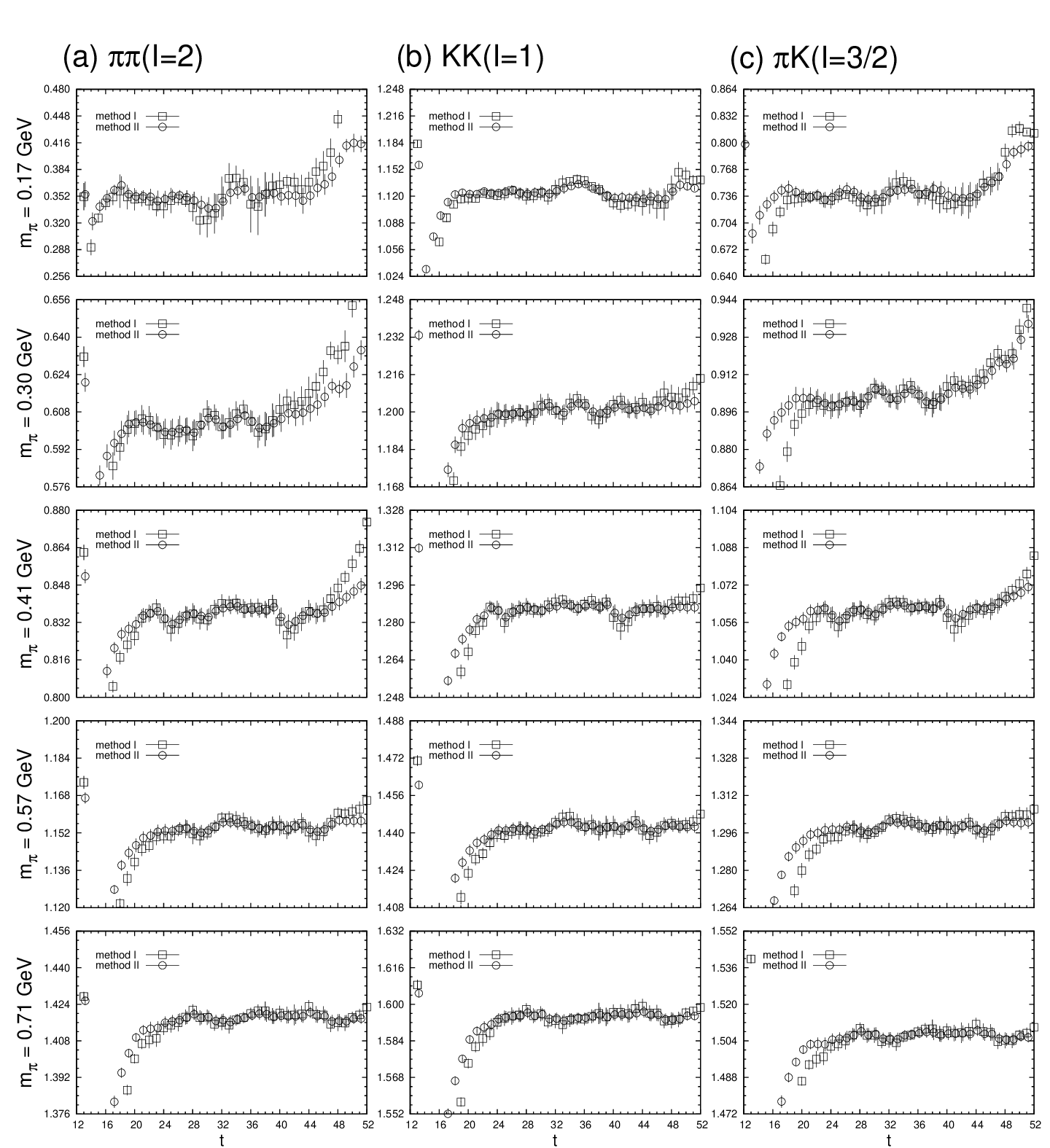}
\caption{
  The effective masses in the physical unit (GeV) for 
  (a) the             $\pi\pi(I=2  )$, 
  (b) the             $ K  K (I=1  )$ and 
  (c) the             $\pi K (I=3/2)$ systems 
  at $m_\pi=0.17$, $0.30$, $0.41$, $0.57$ and $0.71$ GeV. 
  The results of the two methods, 
  the method I (squares) and II (circles), are shown.}
\label{fig:effective-mass1}
\end  {center}
\end  {figure}

The effective masses for the repulsive channel, 
$\pi\pi(I=2  )$, 
$ K  K (I=1  )$ and 
$\pi K (I=3/2)$ systems 
are plotted in the columns (a), (b) and (c) 
of Fig. \ref{fig:effective-mass1}, respectively. 
Each row of Fig. \ref{fig:effective-mass1} represents the effective masses 
for $m_\pi=0.17$, $0.30$, $0.41$, $0.57$ and $0.71$ GeV. 
We show the results with the method I by squares and II by circles. 
We observe clear plateaus for all the cases. 
We also find that the two methods 
give the same results of the effective masses. 
This supports that our choice of $t_1=53$ 
is enough large for the dominance of the one particle state 
in the time correlation functions.

\begin{figure}[htbp]
\begin{center}
\includegraphics[height=180mm]{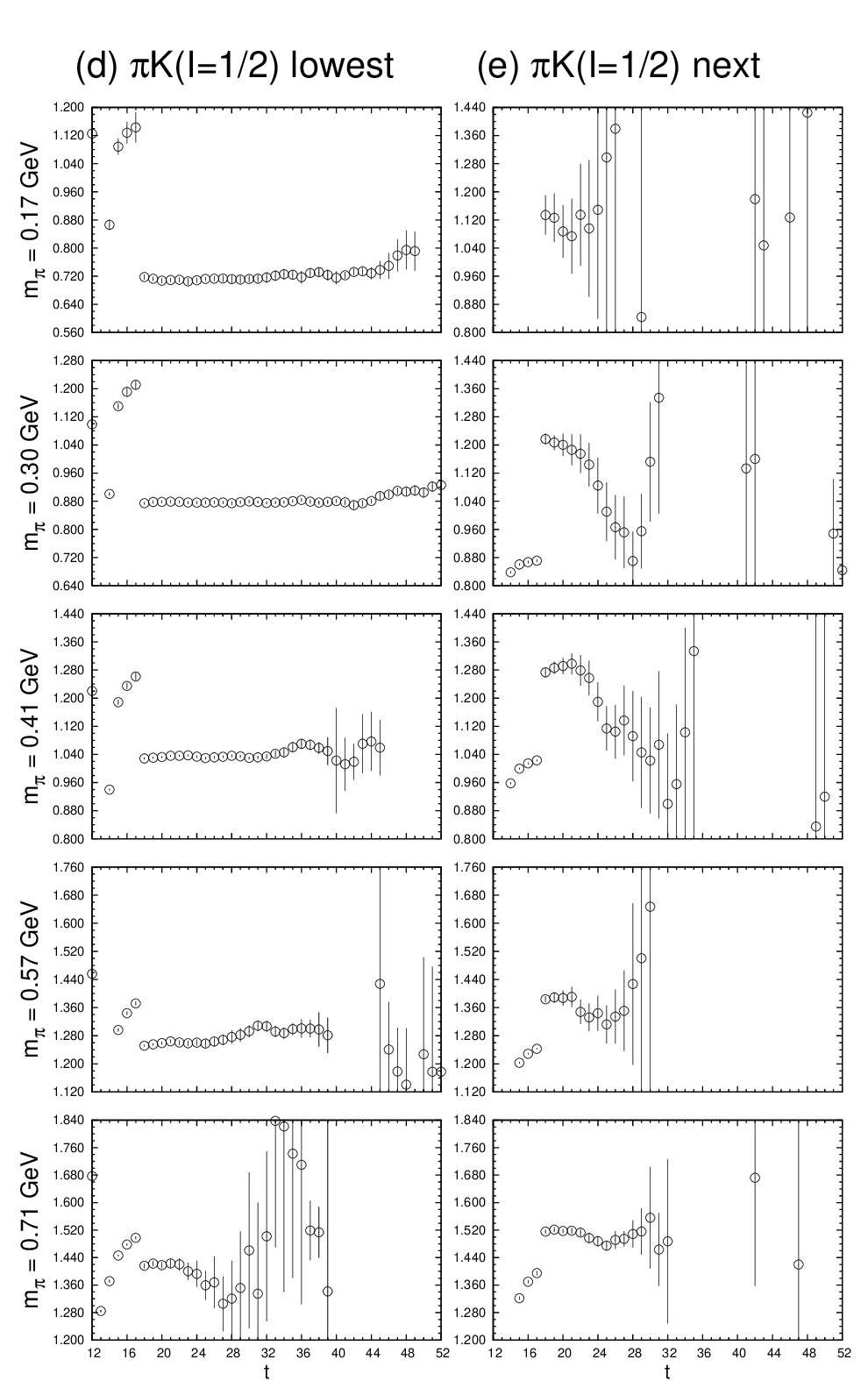}
\caption{
  The effective masses in the physical unit (GeV) for 
  (d) the      lowest $\pi K (I=1/2)$ ($n=0$) and 
  (e) the next-lowest $\pi K (I=1/2)$ ($n=1$) systems 
  at $m_\pi=0.17$, $0.30$, $0.41$, $0.57$ and $0.71$ GeV.} 
\label{fig:effective-mass2}
\end  {center}
\end  {figure}

For the $\pi K (I=1/2)$ system, 
we calculate the effective masses for the eigenvalues 
$\bar{G}_n(t)=\mbox{EV}
 \left[\ [\, G^{(1/2)}(t_R)\ ]^{-1}\,\cdot G^{(1/2)}(t)\,
 \right]_n$ 
in Eq.(\ref{eq:barG})
for 
the      lowest ($n=0$) state and
the next-lowest ($n=1$) state. 
They are plotted in the columns 
(d) and (e) of Fig. \ref{fig:effective-mass2}. 
One sees that 
the effective mass of the $n=0$ state shows a clear plateau 
in the small $m_\pi$ region ($m_\pi=0.17-0.57$ GeV), 
while it does only a short plateau 
at the large $m_\pi$ ($m_\pi=0.71$ GeV). 
The reason will be discussed in Sec.
\ref{ssec:Scattering_length_for_attractive_channel}. 

%
%---------------------------------
%---------------------------------

%
%---------------------------------
%---------------------------------

%
% @@@ =======================================================================
%
\subsection{Scattering length for repulsive channels ($\pi\pi(I=2)$, $KK(I=1)$ and $\pi K(I=3/2)$)}
\label{ssec:Scattering_length_for_repulsive_channels}

For the $\pi\pi(I=2)$, $KK(I=1)$ and $\pi K (I=3/2)$ systems, 
we extract the energy of the lowest state 
by a single exponential fit for 
the time correlation functions, 
$G^{(2)}(t)$, $G^{(1)}(t)$ and $G^{(3/2)}(t)$
in Eqs.(\ref{eqn:TCor_pp2}), (\ref{eqn:TCor_kk1}) and (\ref{eqn:TCor_pk3}). 
As shown in Sec.\ref{ssec:Time_correlation_functions_and_effective_masses}, 
the effective masses of the time correlation functions 
obtained by the methods I and II give the consistent results. 
Thus, we average over the energies 
extracted from these two time correlation functions. 
In Table \ref{tbl:Summary_pp2},
         \ref{tbl:Summary_kk1} and
         \ref{tbl:Summary_pk3}, 
we tabulate the fit range, 
the energy $E$ and the scattering momentum $k$. 
We evaluate the scattering phase shift $\delta_0(k)$ 
by substituting $k$ into Eq.(\ref{eqn:Luschers_formula}), 
which is also tabulate in the tables. 
In all the cases, $\tan\delta_0$ is negative, 
so that the interaction is repulsive. 
If the interaction between two particles is not strong, 
then the scattering momentum $|k^2|$ takes small value 
and $\tan\delta_0(k)/k$ can be expanded in terms of $k^2$ as 
\begin{equation}
  [\, \tan\delta_0(k)/k\, ]^{-1}
  =
    \frac{1}{a_0}
  + \frac{1}{2}r_{\rm eff}k^2
  + {\cal O}(k^4)
\ ,
\label{eqn:eff_range_exp}
\end  {equation}
where 
$a_0$ is the scattering length and $r_{\rm eff}$ is the effective range. 
In the following, 
we assume that 
${\cal O}(k^2)$ and the higher terms can be neglected 
in (\ref{eqn:eff_range_exp}) 
at all $m_\pi$ for the repulsive channel, and 
we regard the first term of (\ref{eqn:eff_range_exp}) 
as the inverse of the scattering length. 
%
%
%----------------------
%
\begin{table}[htbp]
\begin{center}
\begin{tabular}{llllccccc}
\hline
\hline
$\kappa_{ud}$ & & $\quad$ & & 
$0.13781$     &  $0.13770$     &  
$0.13754$     &  $0.13727$     &  $0.13700$     \\
$m_\pi$ & [GeV] & & & 
$0.1661(58)$  &  $0.2973(23)$  &  
$0.4144(16)$  &  $0.5746(13)$  &  $0.7069(12)$  \\
\hline
fit range & & & & 
$22-42$       &  $23-43$       &  
$24-44$       &  $25-46$       &  $25-46$       \\
$E^{\rm free}$ & [GeV] & & & 
$0.332 (12)$  &  $0.5947(47)$  &  
$0.8288(32)$  &  $1.1492(26)$  &  $1.4137(25)$  \\
$E$ & [GeV]            & & & 
$0.347 (10)$  &  $0.6024(49)$  &  
$0.8357(33)$  &  $1.1543(27)$  &  $1.4180(25)$  \\
$k^2$          & [GeV${}^2$] & & & 
$0.00256(25)$ &  $0.00229(23)$ &  
$0.00285(18)$ &  $0.00292(18)$ &  $0.00302(11)$ \\
$[\, \tan\delta_0(k)/k\, ]^{-1}$ & [GeV] & & & 
$-1.78 (16)$  &  $-1.96 (18)$  &  
$-1.611(89)$  &  $-1.574(87)$  &  $-1.531(48)$  \\
$[\, \tan\delta_0(k)/k\, ]$ & [fm] & & & 
$-0.1113(97)$ &  $-0.1010(92)$ &  
$-0.1225(68)$ &  $-0.1254(70)$ &  $-0.1289(41)$ \\
$[\, \tan\delta_0(k)/k\, ]\cdot m_\pi$ & & & &  
$-0.0936(64)$ &  $-0.152 (14)$ & 
$-0.257 (14)$ &  $-0.365 (21)$ &  $-0.462 (15)$ \\
$\delta_0(k)$ & [deg.] & & & 
$-1.64(22)$   &  $-1.40(20)$   &  
$-1.90(16)$   &  $-1.97(17)$   &  $-2.05(10)$   \\
\hline
\hline
\end  {tabular}
\caption{
  The energy $E$, the scattering momentum $k$ 
  and the scattering phase shift $\delta_0(k)$ 
  for the lowest state of the $S$-wave $\pi\pi (I=2)$ system. 
  The fit range for the extraction of the energy $E$ 
  from the time correlation are also tabulated. 
  $E^{\rm free}=2 m_\pi$ is also shown for a guide. 
  }
\label{tbl:Summary_pp2}
\end  {center}
\end  {table}
%
%----------------------
%

%
%----------------------
%
\begin{table}[htbp]
\begin{center}
\begin{tabular}{llllccccc}
\hline
\hline
$\kappa_{ud}$ & & $\quad$ & & 
$0.13781$     &  $0.13770$     &  
$0.13754$     &  $0.13727$     &  $0.13700$     \\
$m_\pi$ & [GeV] & & & 
$0.1661(58)$  &  $0.2973(23)$  &  
$0.4144(16)$  &  $0.5746(13)$  &  $0.7069(12)$  \\
\hline
fit range & & & & 
$22-42$       &  $23-43$       &  
$24-44$       &  $25-46$       &  $25-46$       \\
$E^{\rm free}$ & [GeV] & & & 
$1.1188(46)$  &  $1.1950(35)$  &  
$1.2802(26)$  &  $1.4381(24)$  &  $1.5905(24)$  \\
$E$ & [GeV] & & & 
$1.1256(43)$  &  $1.2003(35)$  &  
$1.2858(25)$  &  $1.4423(25)$  &  $1.5944(24)$  \\
$k^2$          & [GeV${}^2$] & & & 
$0.00382(71)$ &  $0.00321(20)$ & 
$0.00362(22)$ &  $0.00303(32)$ &  $0.00307(22)$ \\
$[\, \tan\delta_0(k)/k\, ]^{-1}$ & [GeV] & & & 
$-1.26 (20)$  &  $-1.450(80)$  &  
$-1.306(70)$  &  $-1.53 (14)$  &  $-1.510(94)$  \\
$[\, \tan\delta_0(k)/k\, ]$ & [fm] & & & 
$-0.158 (25)$ &  $-0.1361(75)$ & 
$-0.1511(80)$ &  $-0.130 (12)$ &  $-0.1308(81)$ \\
$[\, \tan\delta_0(k)/k\, ]\cdot m_K$ & & & & 
$-0.448(71)$  &  $-0.412(23)$  &  
$-0.490(26)$  &  $-0.471(44)$  &  $-0.527(33)$  \\
$[\, \tan\delta_0(k)/k\, ]\cdot m_\pi$ & & & & 
$-0.133(19)$  &  $-0.205(11)$  &  
$-0.317(17)$  &  $-0.377(35)$  &  $-0.468(29)$  \\
$\delta_0(k)$ & [deg.] & & & 
$-2.84(71)$   &  $-2.24(19)$   &  
$-2.64(22)$   &  $-2.07(30)$   &  $-2.10(21)$   \\
\hline
\hline
\end  {tabular}
\caption{The same as Table~\ref{tbl:Summary_pp2}
  for the $S$-wave $KK(I=1)$ system ($E^{\rm free}=2 m_K$). }
\label{tbl:Summary_kk1}
\end  {center}
\end  {table}
%
%----------------------
%

%
%----------------------
%
\begin{table}[htbp]
\begin{center}
\begin{tabular}{llllccccc}
\hline
\hline
$\kappa_{ud}$ & & $\quad$ & & 
$0.13781$     &  $0.13770$     &  
$0.13754$     &  $0.13727$     &  $0.13700$     \\
$m_\pi$ & [GeV] & & & 
$0.1661(58)$  &  $0.2973(23)$  &  
$0.4144(16)$  &  $0.5746(13)$  &  $0.7069(12)$  \\
\hline
fit range & & & & 
$22-42$       &  $23-43$       &  
$24-44$       &  $25-46$       &  $25-46$       \\
$E^{\rm free}$ & [GeV] & & & 
$0.7261(82)$  &  $0.8949(38)$  &  
$1.0545(28)$  &  $1.2937(25)$  &  $1.5021(24)$  \\
$E$ & [GeV] & & & 
$0.7371(73)$  &  $0.9019(39)$  &  
$1.0609(29)$  &  $1.2985(26)$  &  $1.5062(25)$  \\
$k^2$          & [GeV${}^2$] & & & 
$0.00302(39)$ &  $0.00281(14)$ &  
$0.00320(19)$ &  $0.00307(20)$ &  $0.00308(13)$ \\
$[\, \tan\delta_0(k)/k\, ]^{-1}$ & [GeV] & & & 
$-1.54 (18)$  &  $-1.629(70)$  &  
$-1.453(78)$  &  $-1.509(86)$  &  $-1.503(54)$  \\
$[\, \tan\delta_0(k)/k\, ]$ & [fm] & & & 
$-0.129 (15)$ &  $-0.1212(52)$ &  
$-0.1359(72)$ &  $-0.1309(75)$ &  $-0.1313(47)$ \\
$[\, \tan\delta_0(k)/k\, ]\cdot \mu_{\pi K}$ & & & & 
$-0.0838(92)$ &  $-0.1219(54)$ &  
$-0.1733(93)$ &  $-0.212 (12)$ &  $-0.2491(91)$  \\
$[\, \tan\delta_0(k)/k\, ]\cdot m_\pi$ & & & & 
$-0.108 (12)$ &  $-0.1826(82)$ &  
$-0.285 (15)$ &  $-0.381 (22)$ &  $-0.471 (17)$  \\
$\delta_0(k)$ & [deg.] & & & 
$-2.06(36)$   &  $-1.87(12)$   &  
$-2.23(19)$   &  $-2.10(19)$   &  $-2.12(12)$   \\
\hline
\hline
\end  {tabular}
\caption{The same as Table~\ref{tbl:Summary_pp2}
  for the $S$-wave $\pi K(I=3/2)$ system ($E^{\rm free}=m_\pi+m_K$). 
  $\mu_{\pi K}\equiv m_\pi m_K/(m_\pi + m_K)$ 
  is the reduced mass of $\pi$ and $K$. }
\label{tbl:Summary_pk3}
\end  {center}
\end  {table}
%
%----------------------
%

%
% @@@ =======================================================================
%
\subsection{Scattering length for attractive channel ($\pi K(I=1/2)$)}
\label{ssec:Scattering_length_for_attractive_channel}
%
% ----------------------------------------------------------------------
%

\begin{figure}[htbp]
\begin{center}
\includegraphics[width=165mm]{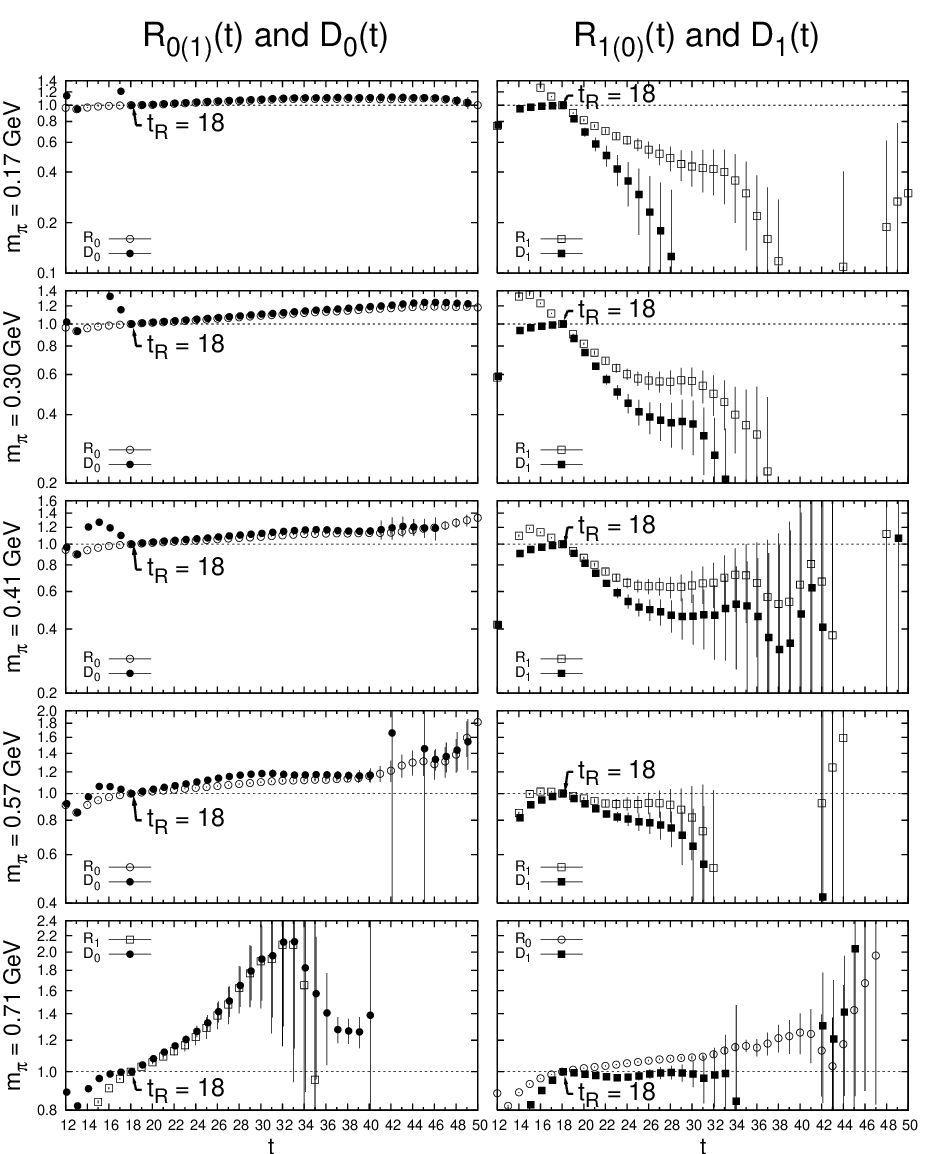}
\caption{
  $R_0(t)$ (open   circles),    $R_1(t)$ (open   squares), 
  $D_0(t)$ (filled circles) and $D_1(t)$ (filled squares) are plotted. 
  Each row represents 
  the results at $m_\pi=0.17$, $0.30$, $0.41$, $0.57$ and $0.71$ GeV, 
  respectively. 
}
\label{fig:corr-ratio}
\end  {center}
\end  {figure}

In order to clearly show 
the contamination from the higher states for $\pi K (I=1/2)$ system, 
we consider the ratios, 
\begin{eqnarray}
  R_i(t)
  &\equiv&
  \frac{G^{(1/2)}_{ii}(t)}{G^{(1/2)}_{ii}(t_R)}
  \cdot
  \left[\ \mbox{e}^{-(m_\pi+m_K)\, (t-t_R)}\ \right]^{-1}
  \quad (i=0,\, 1)
\ ,
\nonumber
\\
  D_n(t) 
  &\equiv&
  \mbox{EV}\left[\ [\, G^{(1/2)}(t_R)\, ]^{-1}\cdot G^{(1/2)}(t)\ \right]_n
  \cdot
  \left[\ \mbox{e}^{-(m_\pi+m_K)\, (t-t_R)}\ \right]^{-1}
  \quad (n=0,\, 1)
\ .
\end  {eqnarray}
In Fig. \ref{fig:corr-ratio}, 
$R_0(t)$ (open   circles),    $R_1(t)$ (open   squares), 
$D_0(t)$ (filled circles) and $D_1(t)$ (filled squares) are plotted. 
Each row of Fig. \ref{fig:corr-ratio} represents 
the results for $m_\pi=0.17$, $0.30$, $0.41$, $0.57$ and $0.71$ GeV, 
respectively. 
Note that for $m_\pi=0.71$ GeV, 
$R_1(t)$ is plotted in the left  panel and 
$R_0(t)$            in the right panel, 
differed from for the other masses.

We find that the difference between $R_0(t)$ and $D_0(t)$ is small 
for $m_\pi=0.17-0.30$ GeV in Fig.\ref{fig:corr-ratio}. 
This means that 
the $\pi K$-type operator ($\Omega^{(1/2)}_0$) 
has a large overlap with the lowest ($n=0$) state 
for the small quark mass. 
On the other hand, at $m_\pi=0.71$ GeV, 
$D_0(t)$ is very different from $R_0(t)$ and is similar to $R_1(t)$. 
This means that 
the operator which has a large overlap with the $n=0$ state 
is the $\bar{s}u$-type operator ($\Omega^{(1/2)}_1$) 
for the large quark mass. 
We can also read out this tendency from the effective masses. 
As we have observed in Fig. \ref{fig:effective-mass2}, 
at $m_\pi=0.71$ GeV, 
the statistical error of the effective mass of the $n=0$ state 
is larger than that of the next-lowest ($n=1$) state. 
This can be attributed to a fact that 
$G_{11}^{(1/2)}(t)$ has a larger statistical error,
and is the dominant contribution to the $n=0$ state.

%
%---------------------------------
%---------------------------------

We show 
the fit range, 
the energy $E$ and the scattering momentum $k$ 
in Table \ref{tbl:Summary_pk1.0} for the      lowest state ($n=0$) 
and      \ref{tbl:Summary_pk1.1} for the next-lowest state ($n=1$). 
We find that 
$k^2$ is negative and the interaction is attractive for the $n=0$ state. 
We evaluate the scattering phase shift $\delta_0(k)$ 
by substituting the $k$ into Eq.(\ref{eqn:Luschers_formula}), 
which are also tabulated in the tables. 
For the $n=1$ state at $m_\pi=0.17$ and $0.71$ GeV, 
$k^2$ gets across 
the divergence points of the function $\sqrt{4\pi}\ g_{00}(k;1)$ 
within the statistical errors and 
the values of $[\, \tan\delta_0(k)/k\, ]^{-1}$ diverge. 
For the $n=1$ state at $m_\pi=0.30$ and $0.41$ GeV, 
$k^2$ gets across 
the zero points of $\sqrt{4\pi}\ g_{00}(k;1)$ 
within the statistical errors and 
the values of $\tan\delta_0(k)/k$ diverge. 
In Table  \ref{tbl:Summary_pk1.1}, these values are omitted.

\begin{figure}[htbp]
\begin{center}
\includegraphics[width=135mm]{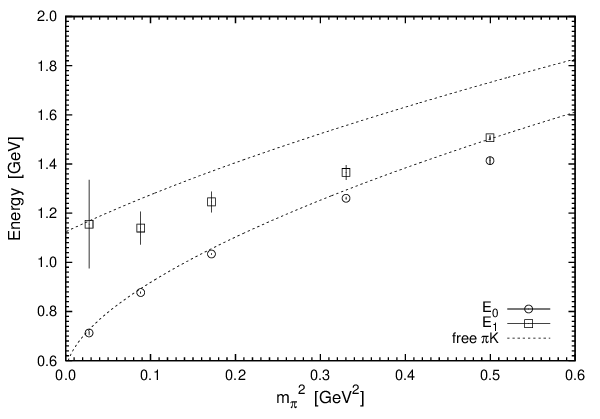}
\caption{The energies of the lowest state (circles)
  and the next-lowest state (squares) for $\pi K (I=1/2)$ system 
  as a function of $m_\pi^2$. 
  The energies of the free $\pi K$ system are also shown by dashed lines. 
}
\label{fig:energy-level.I12}
\end  {center}
\end  {figure}

In Fig.~\ref{fig:energy-level.I12}, 
the energy eigenvalues of the $n=0$ and $n=1$ states 
are plotted as a function of $m_\pi^2$. 
The two dashed lines are the energies of the $n=0$ and $n=1$ states 
for the non-interacting $\pi K$ system, which are given by 
\begin{eqnarray}
  E_0^{\rm free} &=& m_\pi + m_K,                                        \\
  E_1^{\rm free} &=& \sqrt{m_\pi^2+(2\pi/La)^2} + \sqrt{m_K^2+(2\pi/La)^2}. 
\end  {eqnarray}
The continuous values of these for $m_\pi^2$ are given by 
a linear interpolation from the measured value of $m_K^2$. 
From Fig.\ref{fig:energy-level.I12}, we find that 
the energies of the $n=0$ states lie near $E_0^{\rm free}$ 
in the small $m_\pi$ region ($m_\pi=0.17-0.41$ GeV), 
while it is lower than $E_0^{\rm free}$ 
in the largest $m_\pi$ ($m_\pi=0.71$ GeV). 
The energy of the $n=1$ state lies near $E_1^{\rm free}$ 
in the smallest $m_\pi$ ($m_\pi=0.17$ GeV). 
It deviates from $E_1^{\rm free}$, and gets closer 
to $E_0^{\rm free}$ for larger $m_\pi$. 
We note that 
similar features of the $n=0$ state in the scalar meson channel 
have already reported in Refs.~\cite{Scalar01,Scalar02,Scalar03,Scalar04}.

%
%----------------------
%
\begin{table}[htbp]
\begin{center}
\begin{tabular}{llllccccc}
\hline
\hline
$\kappa_{ud}$ & & $\quad$ & & 
$0.13781$     &  $0.13770$     &  
$0.13754$     &  $0.13727$     &  $0.13700$     \\
$m_\pi$ & [GeV] & & & 
$0.1661(58)$  &  $0.2973(23)$  &  
$0.4144(16)$  &  $0.5746(13)$  &  $0.7069(12)$  \\
\hline
fit range & & & & 
$20-42$        &  $20-40$       &  
$20-32$        &  $20-27$       &  $20-27$       \\
$E_0^{\rm free}$ & [GeV] & & & 
$0.7261(82)$   &  $0.8949(38)$   &  
$1.0545(28)$   &  $1.2937(25)$   &  $1.5021(24)$ \\
$E$ & [GeV] & & & 
$0.7126(84)$   &  $0.8772(41)$   &  
$1.0340(27)$   &  $1.2605(57)$   &  $1.413 (15)$ \\
$k^2$          & [GeV${}^2$] & & & 
$-0.00323(59)$ &  $-0.00689(49)$ &  
$-0.01018(78)$ &  $-0.0209 (29)$ &  $-0.064(10)$ \\
$[\, \tan\delta_0(k)/k\, ]^{-1}$ & [GeV] & & & 
$1.07 (23)$    &  $0.380(42)$    &  
$0.184(32)$    &  $-0.038(32)$   &  $-0.239(26)$ \\
$[\, \tan\delta_0(k)/k\, ]$ & [fm] & & & 
$0.188(41)$    &  $0.520(59)$    &  
$1.08 (18)$    &  $-5.9(6.8)$    &  $-0.826(89)$ \\
$[\, \tan\delta_0(k)/k\, ]\cdot\mu_{\pi K}$ & & & & 
$0.122(27)$    &  $0.523(60)$    &  
$1.37 (24)$    &  $-10.(11.)$    &  $-1.57(17)$  \\
$[\, \tan\delta_0(k)/k\, ]\cdot m_\pi$ & & & & 
$0.158(36)$    &  $0.784(89)$    &  
$2.26 (39)$    &  $-17.(20.)$    &  $-2.96(32)$  \\
$\sigma_0(k)$ & [deg.] & & & 
$ 3.12(98)$     &  $12.3(1.8)$   &  
$28.8(5.1)$     &  $-75.(11.)$   &  $-46.68(74)$ \\
\hline
\hline
\end  {tabular}
\caption{The same as Table~\ref{tbl:Summary_pp2}
  for the lowest ($n=0$) state of the $S$-wave $\pi K(I=1/2)$ system 
  ($E_0^{\rm free}=m_\pi+m_K$) except for $\sigma_0(k)$. 
  $\sigma_0(k)$ is a real function defined by the analytic continuation 
  as $\tan\sigma_0/\kappa=\tan\delta_0(k)/k\quad (\kappa^2\equiv -k^2)$. 
  It is noted that $\sigma_0(k)$ is not the physical scattering phase shift.}
\label{tbl:Summary_pk1.0}
\end  {center}
\end  {table}
%
%----------------------
%

%
%----------------------
%
\begin{table}[htbp]
\begin{center}
\begin{tabular}{llllccccc}
\hline
\hline
$\kappa_{ud}$ & & $\quad$ & & 
$0.13781$     &  $0.13770$     &  
$0.13754$     &  $0.13727$     &  $0.13700$     \\
$m_\pi$ & [GeV] & & & 
$0.1661(58)$  &  $0.2973(23)$  &  
$0.4144(16)$  &  $0.5746(13)$  &  $0.7069(12)$  \\
\hline
fit range & & & & 
$20-42$        &  $20-40$       &  
$20-32$        &  $20-27$       &  $20-27$      \\
$E_1^{\rm free}$ & [GeV] & & & 
$1.1679(40)$   &  $1.2600(26)$  &  
$1.3693(21)$   &  $1.5563(21)$  &  $1.7322(21)$ \\
$E$ & [GeV] & & & 
$1.16 (18)$    &  $1.139(67)$   &  
$1.246(42)$    &  $1.366(30)$   &  $1.507(10)$  \\
$k^2$          & [GeV${}^2$] & & & 
$0.182(99)$   &  $0.116(37)$    &  
$0.106(26)$   &  $0.047(20)$    &  $0.0036(78)$ \\
$[\, \tan\delta_0(k)/k\, ]^{-1}$ & [GeV] & & & 
$-$           &  $0.18(29)$     &  
$0.10(15)$    &  $-0.174(93)$   &  $-$           \\
$[\, \tan\delta_0(k)/k\, ]$ & [fm] & & & 
$-$           &  $-$           &  
$-$           &  $-1.18(63)$   &  $-$           \\
$[\, \tan\delta_0(k)/k\, ]\cdot\mu_{\pi K}$ & & & & 
$-$           &  $-$           &  
$-$           &  $-1.9(1.0)$   &  $-$           \\
$[\, \tan\delta_0(k)/k\, ]\cdot m_\pi$ & & & & 
$-$           &  $-$           &  
$-$           &  $-3.4 (1.8)$  &  $-$           \\
$\delta_0(k)$ & [deg.] & & & 
$-$           &  $-$           &  
$-$           &  $-52.(20)$    &  $-$           \\
\hline
\hline
\end  {tabular}
\caption{The same as Table~\ref{tbl:Summary_pp2}
  for the next-lowest ($n=1$) state of the $S$-wave $\pi K(I=1/2)$ system. 
  We take $E_1^{\rm free}=\sqrt{m_\pi^2+p^2}
                         +\sqrt{m_K^2+p^2}$ with $p=2\pi/L$. 
  }
\label{tbl:Summary_pk1.1}
\end  {center}
\end  {table}
%
%----------------------
%

%
%----------------------
%
%----------------------
%

%
%------------------------------------------------------------------
%

\begin{figure}[htbp]
\begin{center}
\includegraphics[width=135mm]{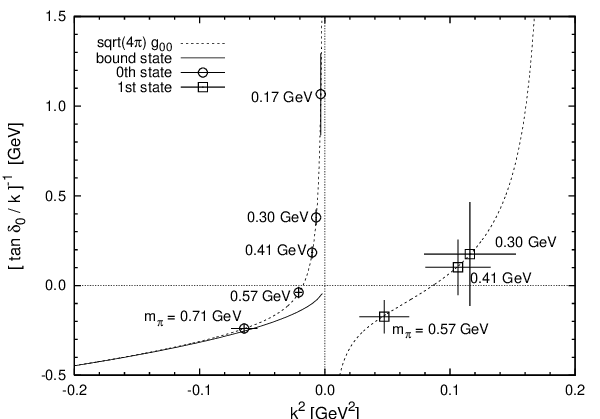}
\caption{$\sqrt{4\pi}g_{00}(k;1)$ (dashed line), and 
  the results of $[\, \tan\delta_0(k)/k\, ]^{-1}$ 
  (open symbols) as a function of $k^2$. 
  A solid line is $\tan\delta_0(k)=-i$ 
  which corresponds to a formation of the bound state.
}
\label{fig:kcotdel.pk1}
\end  {center}
\end  {figure}

In order to more clearly show this phenomena, 
we plot $[\, \tan\delta_0(k)/k\, ]^{-1}$ 
in Fig. \ref{fig:kcotdel.pk1}, 
where the dashed line is the function 
given by the right hand side of Eq.(\ref{eqn:Luschers_formula}). 
A solid line is $\tan\delta_0(k)=-i$. 
The $n=1$ states at $m_\pi=0.17$, $0.71$ GeV are not plotted 
because they across the divergent 
points of $\sqrt{4\pi}\ g_{00}(k;1)$. 
Due to the strong attraction, 
$[\, \tan\delta_0(k)/k\, ]^{-1}$ 
of the $n=0$ state changes the sign, 
and $\tan\delta_0(k)\simeq -i$ at $m_\pi=0.71$ GeV. 
This suggests a bound state formation at $m_\pi=0.71$ GeV.

In the following discussion, we concentrate only on the $n=0$ state, 
because the statistics of the phase shift for the $n=1$ state is not enough 
to analyze the quark-mass dependence and 
obtain the value at physical quark mass. 
Fig. \ref{fig:energy-level.I12} 
and \ref{fig:kcotdel.pk1} suggest that 
the values of the scattering phase shift of $n=0$ state 
for $m_\pi\ge 0.57$ GeV 
might be strongly affected by the existence of the bound state. 
In that case, we need the higher order term of 
$k^2$ in Eq.(\ref{eqn:eff_range_exp}) to obtain the scattering length. 
In the present work, 
we assume that ${\cal O}(k^2)$ terms in Eq.(\ref{eqn:eff_range_exp}) 
can be neglected for $m_\pi\le 0.41$ GeV, 
and we regard $\tan\delta_0(k)/k$ as the scattering length $a_0$. 
%
%----------------------
%
%----------------------
%

%
% @@@ =======================================================================
%
\subsection{Comparison with the previous studies at several quark masses}
\label{ssec:Comparison_with_the_previous_studies_at_several quark masses}

\begin{figure}[htbp]
\begin{center}
\includegraphics[width=160mm]{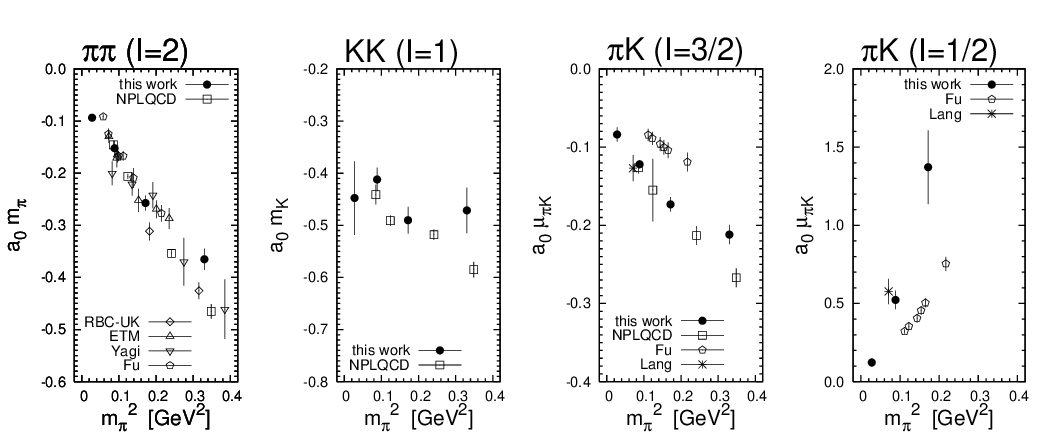}
\caption{Comparison of the present results with the previous lattice studies 
  \cite{NPLQCD.pp2,RBC.pp,ETM.pp2,Yagi.pp2,
        Fu.pp,NPLQCD.KK1,NPLQCD.pK,Fu.pK,Lang.pK}. 
  $a_0^{(2  )} m_\pi     $ for $\pi\pi(I=2  )$, 
  $a_0^{(1  )} m_K       $ for $ K  K (I=1  )$, 
  $a_0^{(3/2)}\mu_{\pi K}$ for $\pi K (I=3/2)$ and  
  $a_0^{(1/2)}\mu_{\pi K}$ for $\pi K (I=1/2)$ are plotted. 
  For Yagi {\it et al.}~\cite{Yagi.pp2}, 
  the finite-volume corrected values are plotted. 
}
\label{fig:comparison}
\end  {center}
\end  {figure}

In this section we compare our results with 
the previous studies 
\cite{NPLQCD.pp2,RBC.pp,ETM.pp2,Yagi.pp2,
      Fu.pp.2,NPLQCD.KK1,NPLQCD.pK,Fu.pK,Lang.pK}. 
In Table \ref{tbl:Setup_comparison}, 
the quark formulation, 
the number of flavor $N_f$, 
the lattice spacing $a$, 
the spatial extent $La$ and 
the pion mass $m_\pi$ 
for the present and previous studies are summarized. 
In Fig.~\ref{fig:comparison}, 
$a_0^{(2  )} m_\pi     $ for $\pi\pi(I=2  )$, 
$a_0^{(1  )} m_K       $ for $ K  K (I=1  )$, 
$a_0^{(3/2)}\mu_{\pi K}$ for $\pi K (I=3/2)$ and  
$a_0^{(1/2)}\mu_{\pi K}$ for $\pi K (I=1/2)$ 
are compared for $m_\pi< 0.63$ GeV, 
where $\mu_{\pi K}\equiv m_\pi m_K/(m_\pi+m_K)$ 
is the reduced mass of $\pi$ and $K$. 
These calculations are performed 
with the different lattice spacings and quark formulations. 
The strange quark mass is set near the physical strange quark mass. 
In the figure, we find that all the results for $\pi\pi$ and $KK$ systems 
are almost consistent. 
Our results for the $\pi K (I=3/2,1/2)$ systems
are reasonably consistent with the ones of 
the NPLQCD Collaboration~\cite{NPLQCD.pK} and 
Lang {\it et al.}~\cite{Lang.pK},
while large discrepancies from Fu's results~\cite{Fu.pK} are found. 
We can consider some possible reasons for the discrepancies, 
{\it e.g.} the discretization error, 
but we need further investigation for a conclusion. 

%
%----------------------
%
\begin{table}[p]
\begin{center}
\begin{tabular}{ccccccccccc}
\hline
\hline
                  & \quad\quad & 
quark formulation & \quad\quad & 
$N_f$             & \quad\quad &
$a$     [fm]      & \quad\quad &
$La$    [fm]      & \quad\quad &
$m_\pi$ [GeV]     \\
\hline
the present work         &  &
improved Wilson          &  & 
$2+1$                    &  &
$0.090$                  &  &
$2.9$                    &  &
$0.17-0.71$              \\
\hline
NPLQCD~\cite{NPLQCD.pp2,NPLQCD.KK1,NPLQCD.pK} &  & 
improved staggered (sea) &  & 
$2+1$                    &  &
$0.125$                  &  &
$2.5$                    &  &
$0.29-0.60$              \\
                         &  & 
$+$ domain-wall (valence) &  & 
                         &  &
                         &  &
                         &  &
                         \\
RBC and UK~\cite{RBC.pp} &  & 
domain-wall              &  & 
$2+1$                    &  &
$0.114$                  &  &
$1.8$                    &  &
$0.43-0.67$              \\
ETM~\cite{ETM.pp2}       &  &
maximally twisted-mass   &  & 
$2$                      &  &
$0.067$                  &  &
$2.1$                    &  &
$0.31$                   \\
                         &  &
                         &  & 
                         &  &
$0.086$                  &  &
$2.1$                    &  &
$0.39-0.49  $            \\
                         &  &
                         &  & 
                         &  &
$0.086$                  &  &
$2.7$                    &  &
$0.27-0.31$              \\
Yagi {\it et al.}~\cite{Yagi.pp2} &  & 
overlap                  &  & 
$2$                      &  &
$0.118$                  &  &
$1.9$                    &  &
$0.29-0.75$              \\
Fu~\cite{Fu.pp.2}        &  & 
improved staggered       &  & 
$2+1$                    &  &
$0.12$                   &  &
$2.7$                    &  &
$0.24$                   \\
                         &  &
                         &  &
                         &  &
                         &  &
$2.4$                    &  &
$0.32-0.37$              \\
                         &  &
                         &  &
                         &  &
$0.15$                   &  &
$3.0$                    &  &
$0.24$                   \\
                         &  &
                         &  &
                         &  &
                         &  &
$2.5$                    &  &
$0.33-0.46$              \\
Fu~\cite{Fu.pK}          &  & 
improved staggered       &  & 
$2+1$                    &  &
$0.15$                   &  &
$2.5$                    &  &
$0.33-0.46$              \\
Lang {\it et al.}~\cite{Lang.pK} &  & 
improved Wilson          &  & 
$2$                      &  &
$0.124$                  &  &
$2.0$                    &  &
$0.27$                   \\
\hline
\hline
\end  {tabular}
\caption{
  The quark formulation, 
  the number of flavor $N_f$, 
  the lattice spacing $a$, 
  the spatial extent $La$ and 
  the pion-mass range 
  of the present and previous studies. 
}
\label{tbl:Setup_comparison}
\end  {center}
\end  {table}
%
%----------------------
%

%
%----------------------
%
%----------------------
%

%
% @@ ========================================================================
%
\section  {Chiral analysis}
\label{sec:Chiral_analysis}
%

%
% @@@ =======================================================================
%

\subsection{Chiral analysis with ${\cal O}(p^4)$ ChPT}
\label{ssec:Chiral_analysis_with_op4_chpt}

In this section, 
we investigate the quark mass dependence of the scattering lengths 
to evaluate the value at the physical quark mass. 
For this purpose, first, 
we consider the ChPT formulae in the ${\cal O}(p^4)$ 
given in Refs.~\cite{Gasser:1984gg,Bernard:1990kw,Chen:2006wf}. 
The scattering lengths of the 
$\pi\pi(I=2  )$, 
$ K  K (I=1  )$, 
$\pi K (I=3/2)$ and 
$\pi K (I=1/2)$ systems 
can be written by 
\begin{eqnarray}
  a_0^{(2)} m_\pi
  & = &
  \frac{m_\pi^2}{16\pi f_\pi^2}
  \left[\
  - 1
  + \frac{16}{f_\pi^2}
    \left[\
              m_\pi^2    \cdot L^\prime
      - \frac{m_\pi^2}{2}\cdot L_5
      + \zeta^{(2)}\
    \right]\
  \right]\ ,
\label{eqn:su3chpt-p4-dcp1-a0pp2}
\\
  a_0^{(1)} m_K
  & = &
  \frac{m_K^2}{16\pi f_K^2}
  \left[\
  - 1
  + \frac{16}{f_K^2}
    \left[\
              m_K^2    \cdot L^\prime
      - \frac{m_K^2}{2}\cdot L_5
      + \zeta^{(1)}\
    \right]\
  \right]\ ,
\label{eqn:su3chpt-p4-dcp1-a0kk1}
\\
  a_0^{(3/2)} \mu_{\pi K}
  & = &
  \frac{\mu_{\pi K}^2}{8\pi f_\pi f_K}
  \left[\
  - 1
  + \frac{16}{f_\pi f_K}
    \left[\
              m_\pi     m_K      \cdot L^\prime
      - \frac{m_\pi^2 + m_K^2}{4}\cdot L_5
      + \zeta^{(3/2)}\
    \right]\
  \right]\ ,
\label{eqn:su3chpt-p4-dcp1-a0pk3}
\\
  a_0^{(1/2)} \mu_{\pi K}
  & = &
  \frac{\mu_{\pi K}^2}{8\pi f_\pi f_K}
  \left[\
    2
  + \frac{16}{f_\pi f_K}
    \left[\
                 m_\pi     m_K       \cdot L^\prime
      + 2 \frac{ m_\pi^2 + m_K^2}{4} \cdot L_5
      + \zeta^{(1/2)}\
    \right]\ 
  \right]\ ,
\label{eqn:su3chpt-p4-dcp1-a0pk1}
\end  {eqnarray}
where 
the formulae are written by 
${\cal O}(p^4)$ values of the masses of the NG bosons ($m_\pi$ and $m_K$) and 
the decay constants ($f_\pi$ and $f_K$), 
which are not the parameter of ChPT and depend on the quark masses. 
The normalization of $f_\pi=0.092$ GeV at the physical point is adopted. 
The constants $L_5$ and 
\begin{equation}
  L^\prime\equiv 2L_1+2L_2+L_3-2L_4-L_5/2+2L_6+L_8
\end{equation}
are the low energy constants (LECs) defined in Ref.~\cite{Gasser:1984gg}
at a renormalization scale $\mu$. 
In the present work, we adopt $\mu=0.770$ GeV. 
$\zeta^{(2),(1),(3/2),(1/2)}$ are known functions 
with chiral logarithmic terms, 
which are given in Appendix \ref{app:op4_su3_wchpt_formulae}.

In the chiral analysis, 
we fit our results of the scattering length
with the ChPT formulae in Eqs.~(\ref{eqn:su3chpt-p4-dcp1-a0pp2})--(\ref{eqn:su3chpt-p4-dcp1-a0pk1})
for all the channels simultaneously,
where 
the values of $m_\pi$, $m_K$, $f_\pi$, and $f_K$ in the formulae 
are fixed to the measured values by the lattice calculations at each quark mass.
The free parameters in the fitting are 
the LECs ($L_5$ and $L^\prime$).

For the other fitting procedure, 
we rewrite the ChPT formulae in terms of the the quark mass
and the decay constant at $m_{\rm q}=0$, $F$, and fit our results 
with the formulae,
regarding the LECs ($L_5$ and $L^\prime$) and $F$ as free parameter 
of the fitting.
In this case the measured values of the decay constants $f_\pi$ and $f_K$ are not used. 
However, it was shown that using the measured values $f_\pi$
significantly improve the convergence of the chiral expansion 
in the studies of the $\pi\pi(I=2)$ scattering lengths 
in Ref.\cite{NPLQCD.pp2,ETM.pp2}. 
Motivated from these studies, 
we adopt the chiral analysis with the measured values in the present work.

Before showing results of the chiral analysis, 
we explain the decay constant used in the present work. 
The values of $f_\pi$ and $f_K$ in the 
same lattice setup have been calculated in Ref.~\cite{Aoki:2008sm}. 
They are defined in the normalization 
with $f_\pi=0.130$ GeV at the physical point 
and evaluated with the perturbative renormalization factor $Z_A^{\rm P}$. 
In the present work, 
we convert them to the values in the normalization with $f_\pi=0.092$ GeV, 
and also to the values 
evaluated with the non-perturbative renormalization factor $Z_A^{\rm NP}$ 
by multiplying $Z_A^{\rm NP}/(u_0 Z_A^{\rm P})$. 
Here, 
$u_0\equiv P^{1/4}$ is the correction factor 
of the tadpole improved renormalization 
with plaquette value $P$ and takes $u_0=0.86968135$
(Ref.~\cite{Aoki:2010wm}). 
The renormalization factors are given by 
$Z_A^{\rm P}=0.94279347$ in 
Refs.~\cite{Aoki:1998ar,Taniguchi:1998pf,Aoki:1998qd} and 
$Z_A^{\rm NP}=0.781(20)$ in Ref.~\cite{Aoki:2010wm}. 
Converted values of $f_\pi$ and $f_K$ used in the present work 
are listed in Table~\ref{tbl:fpi_fkn}. 
Here, the systematic uncertainty 
for the determination of the renormalization factor $Z_A^{\rm NP}$ 
is added to the statistical errors in quadrature.

%
%---------------------------------
\begin{table}[htbp]
\begin{center}
\begin{tabular}{ccccccccccc}
\hline
\hline
%$m_\pi$ [GeV] & $\quad$ & 
%$0.1661(58)$  & $\quad$ & 
%$0.2973(23)$  & $\quad$ & 
%$0.4144(16)$  & $\quad$ & 
%$0.5746(13)$  & $\quad$ & 
%$0.7069(12)$  \\
$m_\pi$ [GeV] & $\quad$ & 
$0.17$        & $\quad$ & 
$0.30$        & $\quad$ & 
$0.41$        & $\quad$ & 
$0.57$        & $\quad$ & 
$0.71$         \\
\hline
$f_\pi$ [GeV] & & 
$0.0969(57)$  & & 
$0.1030(29)$  & & 
$0.1105(29)$  & & 
$0.1260(42)$  & & 
$0.1327(38)$  \\
$f_K$   [GeV] & & 
$0.1148(35)$  & & 
$0.1195(32)$  & & 
$0.1246(33)$  & & 
$0.1353(41)$  & & 
$0.1392(40)$  \\
\hline
\hline
\end  {tabular}
\caption{
  The values of $f_\pi$ and $f_K$ used in the present work. }
\label{tbl:fpi_fkn}
\end  {center}
\end  {table}
%
%---------------------------------
%

Here, we show the results of the analysis with the ChPT formulae of 
Eqs.(\ref{eqn:su3chpt-p4-dcp1-a0pp2}),
    (\ref{eqn:su3chpt-p4-dcp1-a0kk1}),
    (\ref{eqn:su3chpt-p4-dcp1-a0pk3}) and
    (\ref{eqn:su3chpt-p4-dcp1-a0pk1}). 
In the fitting, 
correlations among the scattering lengths for the the different channels 
are taken into account by the covariance matrix among them. 
The statistical errors of the fitting results of LECs 
are evaluated by the jackknife method. 
The errors of $f_\pi$ and $f_K$ are not included. 
The systematic error from the uncertainty 
for $f_\pi$ and $f_K$ is discussed 
in Sec. \ref{sssec:Uncertainty_of_fpi_and_fkn}.

It was found in Ref.~\cite{Aoki:2008sm} that
the formulae of ${\cal O}(p^4)$ ChPT describe the quark-mass dependence
for $m_\pi$, $m_K$ and $f_\pi$ in $m_\pi\le 0.41$ GeV, 
and $f_K$                      in $m_\pi\le 0.30$ GeV well.
From this, we consider that the formula of ${\cal O}(p^4)$ ChPT 
can be safely applied to our scattering length in these mass ranges. 
In the present work, for the repulsive channels, 
we analyze the following data, 
\begin{eqnarray}
  a_0^{(2  )} m_\pi
  & &
  \mbox{for}\ \ m_\pi=0.17,\ 0.30,\ 0.41\ \mbox{GeV}\ ,
\nonumber  \\
  a_0^{(1  )} m_K
  & &
  \mbox{for}\ \ m_\pi=0.17,\ 0.30\ \mbox{GeV}\ ,
\nonumber  \\
  a_0^{(3/2)} \mu_{\pi K}
  & &
  \mbox{for}\ \ m_\pi=0.17,\ 0.30\ \mbox{GeV}\ .
\label{eqn:data_set_for_replsive_channels}
\end  {eqnarray}
For the $\pi K (I=1/2)$ system, in the continuum theory, 
it is known that the convergence of the ChPT is not good 
compared with those for the repulsive channel. 
Thus, we need to test the convergence of the ChPT formula in this channel. 
In the present work, 
we consider the following three data sets of $a_0^{(1/2)}\mu_{\pi K}$ 
with the data in Eq.(\ref{eqn:data_set_for_replsive_channels}) 
for the chiral analysis, and investigate the stability of the fitting. 
\begin{eqnarray}
  {\rm data\ set\ A} &:& {\rm not\ data}\ ,
\nonumber
\\
  {\rm data\ set\ B} &:& m_\pi= 0.17\ {\rm GeV}\ ,
\nonumber
\\
  {\rm data\ set\ C} &:& m_\pi=0.17,\ 0.30\ {\rm GeV}\ ,
\end  {eqnarray}
where these data sets are called data set A, B and C.

\begin{figure}[htbp]
\begin{center}
\includegraphics[width=148mm]{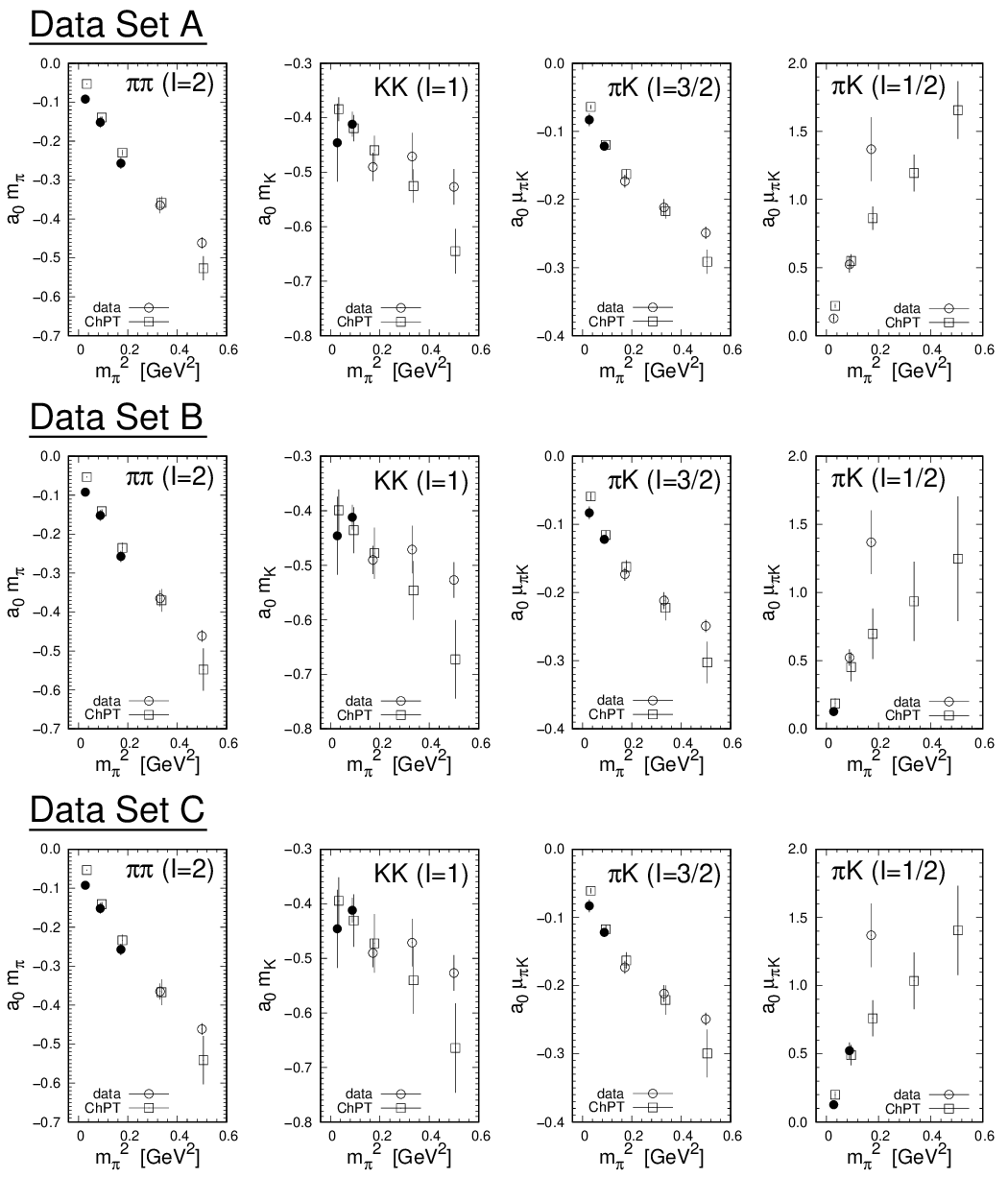}
\caption{
  Fitting results of the ${\cal O}(p^4)$ ChPT fit 
  with the data set A, B and C. 
  The data are represented by circles, 
  and those used in the fit by filled symbols. 
  The fitting results by the ChPT formulae are denoted by squares.
}
\label{fig:su3chpt-p4-a0mm.www0}
\end  {center}
\end  {figure}

\begin{figure}[htbp]
\begin{center}
\includegraphics[width=148mm]{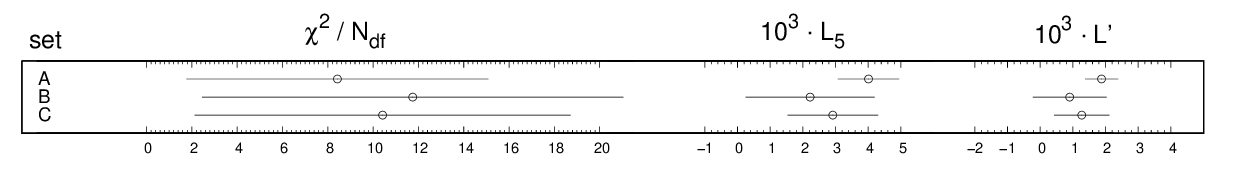}
\caption{
  $\chi^2/N_{\rm df}$ and
  LECs ($10^3\cdot L_5     $ and 
        $10^3\cdot L^\prime$)
  determined in the ${\cal O}(p^4)$ ChPT fit 
  with the data sets A, B and C. 
}
\label{fig:su3chpt-p4-summary}
\end  {center}
\end  {figure}

In Fig.\ref{fig:su3chpt-p4-a0mm.www0}, 
we plot the fitting results of the ChPT formulae 
with the data sets A, B and C. 
In all the cases, the ChPT formulae reproduce the data for 
$a_0^{(2  )} m_\pi$ in $m_\pi=0.30, 0.41$ GeV, 
$a_0^{(1  )} m_K  $, 
$a_0^{(3/2)} \mu_{\pi K}$ 
and 
$a_0^{(1/2)} \mu_{\pi K}$ at $m_\pi=0.30$ GeV well. 
At $m_\pi=0.17$ GeV, however, 
the fitting results for some channels 
are not consistent with the data points. 
The deviation between the data and the fitting results 
can be also seen in the values of $\chi^2/N_{\rm df}$ 
($N_{\rm df}$ is the degrees of freedom in the fit), 
which are plotted in Fig.\ref{fig:su3chpt-p4-summary} 
together with results of 
LECs ($10^3\cdot L_5$ and 
      $10^3\cdot L^\prime$). 
For each data set, 
$\chi^2/N_{\rm df}$ takes huge number, ${\cal O}(10)$. 
This shows that the fitting with the ${\cal O}(p^4)$ ChPT formulae 
in Eqs.~(\ref{eqn:su3chpt-p4-dcp1-a0pp2})--(\ref{eqn:su3chpt-p4-dcp1-a0pk1}) 
does not work for our results of the scattering length. 

%
%----------------------
%
%----------------------
%

%
%----------------------
%
%----------------------
%

%
% @@@ =======================================================================
%

\subsection{Chiral analysis with ${\cal O}(p^4)$ WChPT}
\label{ssec:Chiral_analysis_with_op4_wchpt}

The scattering length vanishes in the chiral limit 
due to the chiral symmetry. 
But, for the Wilson fermion, 
it does not vanish 
due to the explicit chiral symmetry breaking from the Wilson term. 
We consider that an effect of this symmetry breaking 
causes the discrepancy between the our data and the formulae of ChPT. 
In order to investigate this, 
we need to consider the ChPT 
including the effect of the explicit chiral symmetry breaking, 
which has been proposed in Refs. 
\cite{Sharpe:1998xm,Rupak:2002sm,Aoki:2003yv,
      Bar:2003mh,Sharpe:2004ny,Aoki:2005mb,Hansen:2011mc}, 
and usually called the WChPT. 
The WChPT formula for the $a_0^{(2)} m_\pi$ has been given 
by Ref.~\cite{Aoki:2008gy} in the case of $N_{\rm f}=2$. 
Here, we extend it to the WChPT formulae 
for the other scattering systems in the case of $N_{\rm f}=2+1$.

When we apply the WChPT to an chiral analysis, 
we choose an appropriate order counting rule for 
the quark mass $m_{\rm q}$ and the lattice cutoff $a$ 
for our lattice data. 
Our calculations are done 
with the non-perturbatively ${\cal O}(a)$-improved theory, 
thus we treat only the terms higher than ${\cal O}(a^2)$ 
in the lagrangian. 
In the present work, 
we adopt the following counting rule, 
\begin{eqnarray}
  {\rm \bf CR1}                   & & \nonumber  \\
  {\rm  LO}  & : &  p^2,\ m_{\rm q}   \nonumber  \\
  {\rm NLO}  & : &  p^4,\ p^2 m_{\rm q},\ m_{\rm q}^2,\ a^2\ .
\label{eqn:counting-rule1}
\end  {eqnarray}
In the following, we call this counting rule CR1. 
A dependence of the choice of the counting rule 
will be discussed in the next section.

In this counting rule, 
the scattering lengths for the 
$\pi\pi(I=2  )$, 
$ K  K (I=1  )$, 
$\pi K (I=3/2)$ and 
$\pi K (I=1/2)$ systems are given by 
\begin{eqnarray}
  a_0^{(2)} m_\pi 
  & = &
  a_0^{(2)} m_\pi\, 
  |_{\, {}_{\rm  ChPT}}
  - 
  \frac{c_2 a^2}{16\pi f_\pi^2}\ ,
\label{eqn:su3wchpt-p4-dcp1-cr1-a0pp2}
\\
  a_0^{(1)} m_K 
  & = &
  a_0^{(1)} m_K\, 
  |_{\, {}_{\rm  ChPT}}
  - 
  \frac{c_2 a^2}{16\pi f_K^2}\ ,
\label{eqn:su3wchpt-p4-dcp1-cr1-a0kk1}
\\
  a_0^{(3/2)} \mu_{\pi K} 
  & = &
  a_0^{(3/2)} \mu_{\pi K}\, 
  |_{\, {}_{\rm  ChPT}}
  - 
  \frac{c_2 a^2}{8\pi f_\pi f_K}\cdot 
  \frac{\mu_{\pi K}^2}{m_\pi m_K}\ ,
\label{eqn:su3wchpt-p4-dcp1-cr1-a0pk3}
\\
  a_0^{(1/2)} \mu_{\pi K} 
  & = &
  a_0^{(1/2)} \mu_{\pi K}\, 
  |_{\, {}_{\rm  ChPT}}
  - 
  \frac{c_2 a^2}{8\pi f_\pi f_K}\cdot 
  \frac{\mu_{\pi K}^2}{m_\pi m_K}\ ,
\label{eqn:su3wchpt-p4-dcp1-cr1-a0pk1}
\end  {eqnarray}
where 
$a_0^{(2  )} m_\pi     \, |_{\rm ChPT}$, 
$a_0^{(1  )} m_K       \, |_{\rm ChPT}$, 
$a_0^{(3/2)}\mu_{\pi K}\, |_{\rm ChPT}$ and 
$a_0^{(1/2)}\mu_{\pi K}\, |_{\rm ChPT}$ are the ChPT formulae given by 
Eqs.(\ref{eqn:su3chpt-p4-dcp1-a0pp2}),
    (\ref{eqn:su3chpt-p4-dcp1-a0kk1}),
    (\ref{eqn:su3chpt-p4-dcp1-a0pk3}) and
    (\ref{eqn:su3chpt-p4-dcp1-a0pk1}), respectively. 
$c_2$ is a LEC of the WChPT. 
The details of these formulae are discussed in 
Appendix~\ref{app:op4_su3_wchpt_formulae}.

\begin{figure}[htbp]
\begin{center}
\includegraphics[width=148mm]{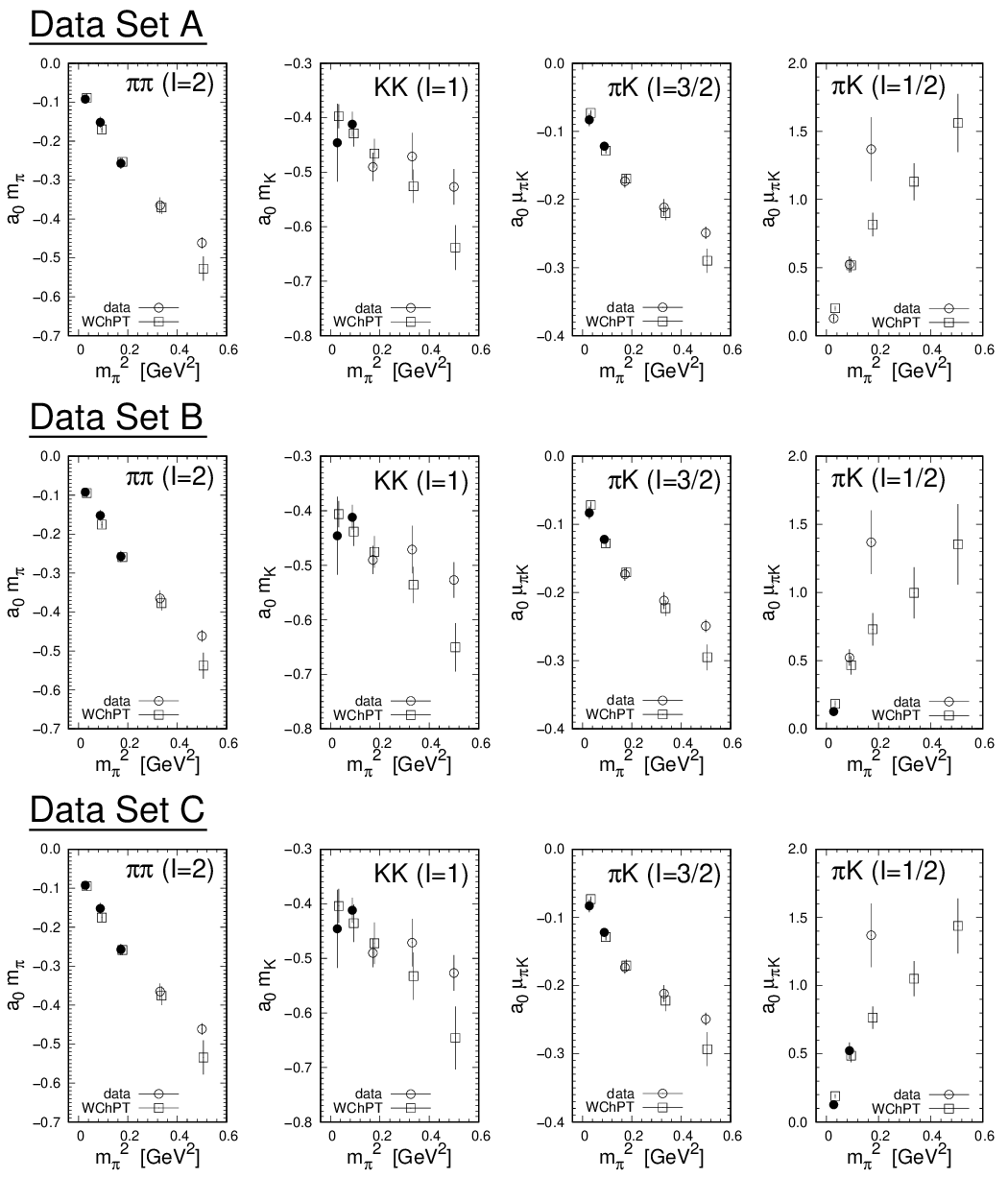}
\caption{
  Fitting results of the ${\cal O}(p^4)$ WChPT fit 
  with the data set A, B and C. 
  The data are represented by circles, 
  and those used in the fit especially done by filled symbols. 
  The fitting results by the WChPT formulae are denoted by squares.
}
\label{fig:su3wchpt-p4-a0mm.w2w0}
\end  {center}
\end  {figure}

\begin{figure}[htbp]
\begin{center}
\includegraphics[width=148mm]{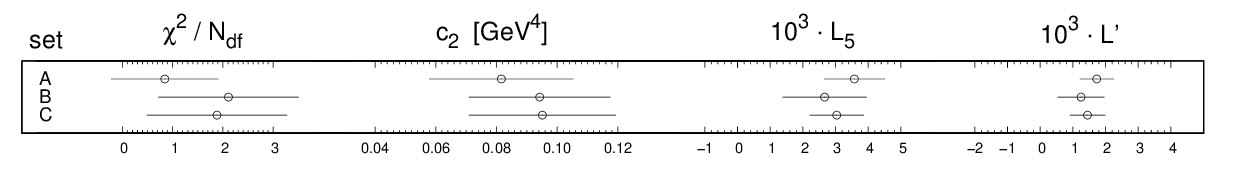}
\caption{
  $\chi^2/N_{\rm df}$ and
  LECs ($c_2$, 
        $10^3\cdot L_5     $ and 
        $10^3\cdot L^\prime$)
  determined in the ${\cal O}(p^4)$ WChPT fit 
  with the data sets A, B and C.
}
\label{fig:su3wchpt-p4-summary}
\end  {center}
\end  {figure}

Like as for the ChPT fit, 
we fix $m_\pi$, $m_K$, $f_\pi$, and $f_K$ in the WChPT formulae 
the measured values 
by the lattice calculations at each quark mass. 
We fit our results with the formulae 
regarding the LECs ($L_5$, $L^\prime$ and $c_2$) as free parameters. 
In Fig.\ref{fig:su3wchpt-p4-a0mm.w2w0}, 
we plot the fitting results of the WChPT formulae 
with the data sets A, B and C. 
We finds that 
the fitting results at $m_\pi=0.17$ GeV are consistent 
with the data points. 
We show more detailed information in Fig.\ref{fig:su3wchpt-p4-summary}, 
where 
$\chi^2/N_{\rm df}$ and 
LECs ($c_2$, 
      $10^3\cdot  L_5     $ and 
      $10^3\cdot  L^\prime$) are given. 
$\chi^2/N_{\rm df}$ is improved 
comparing with the ChPT fitting, 
and takes the reasonable value within the statistical error. 
We find that the fittings 
for three data sets give consistent results. 
This means that the WChPT formula works well
for $a_0^{(2)}$ for $m_\pi \le 0.41\ \mbox{GeV}$, 
$a_0^{(1)}$ and $a_{0}^{(3/2)}$ for $m_\pi \le 0.30\ \mbox{GeV}$,
and $a_{0}^{(1/2)}$ for $m_\pi \le 0.30\ \mbox{GeV}$. 

%
%----------------------
%
%----------------------
%

%
%----------------------
%
%----------------------
%

\begin{figure}[htbp]
\begin{center}
\includegraphics[width=160mm]{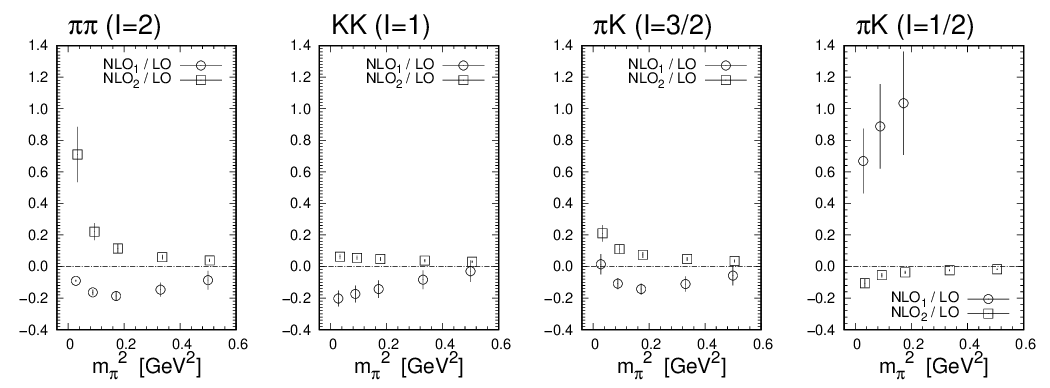}
\caption{The ratio of the next leading terms to the leading term. 
  NLO$_1$/LO (circles) and NLO$_2$ (squares) are represented. 
}
\label{fig:su3wchpt_convergence}
\end  {center}
\end  {figure}

To investigate the convergence of the WChPT formulae, 
we consider the ratio of the next leading terms to the leading term in 
Eqs.(\ref{eqn:su3wchpt-p4-dcp1-cr1-a0pp2}),
    (\ref{eqn:su3wchpt-p4-dcp1-cr1-a0kk1}),
    (\ref{eqn:su3wchpt-p4-dcp1-cr1-a0pk3}) and
    (\ref{eqn:su3wchpt-p4-dcp1-cr1-a0pk1}). 
In Fig.\ref{fig:su3wchpt_convergence}, 
we show 
\begin{equation}
  \frac{{\rm NLO}_1}{\rm LO}
  =
  -\frac{16}{f_\pi^2}
    \left[\
              m_\pi^2    \cdot L^\prime
      - \frac{m_\pi^2}{2}\cdot L_5
      + \zeta^{(2)}\
    \right]
\ ,\quad
  \frac{{\rm NLO}_2}{\rm LO}
  =
  \frac{c_2 a^2}{m_\pi^2}
\ ,
\end  {equation}
\begin{equation}
  \frac{{\rm NLO}_1}{\rm LO}
  =
  -\frac{16}{f_K^2}
    \left[\
              m_K^2    \cdot L^\prime
      - \frac{m_K^2}{2}\cdot L_5
      + \zeta^{(1)}\
    \right]
\ ,\quad
  \frac{{\rm NLO}_2}{\rm LO}
  =
  \frac{c_2 a^2}{m_K^2}
\ ,
\end  {equation}
\begin{equation}
  \frac{{\rm NLO}_1}{\rm LO}
  =
  -\frac{16}{f_\pi f_K}
    \left[\
              m_\pi m_K          \cdot L^\prime
      - \frac{m_\pi^2 + m_K^2}{4}\cdot L_5
      + \zeta^{(3/2)}\
    \right]
\ ,\quad
  \frac{{\rm NLO}_2}{\rm LO}
  =
  \frac{c_2 a^2}{m_\pi m_K}
\ ,
\end  {equation}
\begin{equation}
  \frac{{\rm NLO}_1}{\rm LO}
  =
  \frac{8}{f_\pi f_K}
    \left[\
               m_\pi m_K          \cdot L^\prime
      + 2\frac{m_\pi^2 + m_K^2}{4}\cdot L_5
      + \zeta^{(1/2)}\
    \right]
\ ,\quad
  \frac{{\rm NLO}_2}{\rm LO}
  =
  - 
  \frac{c_2 a^2}{2 m_\pi m_K}
\ ,
\end  {equation}
for the 
$\pi\pi(I=2  )$, 
$ K  K (I=1  )$, 
$\pi K (I=3/2)$ and 
$\pi K (I=1/2)$ systems, respectively. 
In this figure, 
we use the LECs ($L^\prime$, $L_5$ and $c_2$) obtained with the data set B. 
For the repulsive channels, the ratios are at most 20\% 
except for NLO$_2$/LO of $a_0^{(2)} m_\pi$ at $m_\pi=0.17$ GeV. 
The irregular NLO$_2$/LO means that 
the effect of the explicit chiral symmetry breaking from the Wilson term 
cannot be negligible compared to the leading term of the WChPT 
for $a_0^{(2)} m_\pi$ at $m_\pi=0.17$ GeV. 
For the $\pi K(I=1/2)$ system, 
we observe that NLO$_1$/LO is not so small over a wide range of $m_\pi^2$. 
The convergence for the $\pi K(I=1/2)$ system might be disputable 
although the WChPT fit seems to work well 
from the point of view of $\chi^2/N_{\rm df}$. 
However, 
the number of data points is insufficient 
to perform the detailed investigation with ${\cal O}(p^6)$ WChPT fit. 
In the present work, alternatively, 
we discuss rough estimations of the ${\cal O}(p^6)$ contributions 
in Sec.\ref{sssec:higher_order_effects_of_ChPT}.

%
%----------------------
%
%----------------------
%

%
% @@ ========================================================================
%
\section  {Extrapolation to the physical point}
\label{sec:extrapolation_to_the_physical_point}
%

%
% @@@ =======================================================================
%

\subsection{Scattering lengths at the physical point}
\label{ssec:scattering_lengths_at_the_physical_point}

We obtain the scattering length at the physical point 
by using the ${\cal O}(p^4)$ ChPT formulae, 
Eqs.~(\ref{eqn:su3chpt-p4-dcp1-a0pp2})--(\ref{eqn:su3chpt-p4-dcp1-a0pk1}), 
with the LECs ($L_5$ and $L^\prime$) 
obtained from the ${\cal O}(p^4)$ WChPT fit 
in Sec.~\ref{ssec:Chiral_analysis_with_op4_wchpt}. 
Here, at the physical point, 
$m_\pi=0.140$ GeV, 
$m_K  =0.494$ GeV, 
$f_\pi=0.092$ GeV and 
$f_K  =0.110$ GeV. 
The results obtained with the data sets A, B and C are listed 
in Table.\ref{tbl:SL_at_physical_point}.

As mentioned in Sec.\ref{ssec:Chiral_analysis_with_op4_wchpt}, 
three data set give consistent results, 
and thus we adopt the data set B for the standard fit. 
The extrapolated results including the systematic error are summarized as 
\begin{eqnarray}
  a_0^{(2  )} m_\pi      &=& -0.04243(22)(43)  \nonumber \\
  a_0^{(1  )} m_K        &=& -0.312  (17)(31)  \nonumber \\
  a_0^{(3/2)}\mu_{\pi K} &=& -0.0477 (27)(20)  \nonumber \\
  a_0^{(1/2)}\mu_{\pi K} &=&  0.150  (16)(37)\ , 
\label{eqn:SCL_at_phys_point}
\end  {eqnarray}
where the first parenthesis is the statistical error 
and the second parenthesis represents the systematic error 
which is discussed in the following subsections. 

%
%----------------------
%
\begin{table}[p]
\begin{center}
\begin{tabular}{ccccc}
\hline
\hline
data set & \quad\quad & 
A              & B              & C               \\
\hline
$\chi^2/N_{\rm df}$ & & 
$ 0.8(1.1)$    & $ 2.1(1.4)$    & $ 1.9(1.4)$     \\
$c_2$ [GeV${}^4$] & & 
$ 0.082(24)$   & $ 0.094(23)$   & $ 0.095(24)$    \\
$10^3\cdot L_5$ & & 
$ 3.58(93)$    & $ 2.7(1.3)$    & $ 3.04(83)$     \\
$10^3\cdot L^\prime$ & & 
$ 1.73(52)$    & $ 1.25(72)$    & $ 1.45(54)$     \\
\hline
$a_0^{(2  )} m_\pi     $ & & 
$-0.04239(21)$ & $-0.04243(22)$ & $-0.04241(29)$  \\
\hline
$a_0^{(1  )} m_K       $ & & 
$-0.309  (16)$ & $-0.312  (17)$ & $-0.343  (23)$  \\
$a_0^{(1  )} m_\pi     $ & & 
$-0.0874 (45)$ & $-0.0885 (48)$ & $-0.0972 (66)$  \\
\hline
$a_0^{(3/2)}\mu_{\pi K}$ & & 
$-0.0497 (20)$ & $-0.0477 (27)$ & $-0.0485 (15)$  \\
$a_0^{(3/2)} m_\pi     $ & & 
$-0.0638 (26)$ & $-0.0612 (35)$ & $-0.0623 (19)$  \\
\hline
$a_0^{(1/2)}\mu_{\pi K}$ & & 
$ 0.162  (12)$ & $ 0.150  (16)$ & $ 0.155  (11)$  \\
$a_0^{(1/2)} m_\pi     $ & & 
$ 0.208  (15)$ & $ 0.193  (21)$ & $ 0.199  (14)$  \\
\hline
\hline
\end  {tabular}
\caption{
  $\chi^2/N_{\rm df}$ and LECs obtained 
  from the ${\cal O}(p^4)$ $SU(3)$ WChPT fits 
  for the data sets A, B and C. 
  The scattering lengths at the physical point are also shown. 
}
\label{tbl:SL_at_physical_point}
\end  {center}
\end  {table}
%
%----------------------
%

%
% @@@ =======================================================================
%

\subsection{Estimate of systematic errors}
\label{ssec:estimate_of_systematic_errors}

%
% @@@ =======================================================================
%

\subsubsection{Choice of the counting rule for the WChPT}
\label  {sssec:Choice_of_the_counting_rule_for_the_WChPT}

In this section, we investigate the 
dependence of the choice 
of the order counting rule for the results of the chiral analysis. 
Here, we consider another counting rule (``counting rule 2''(CR2)), 
\begin{eqnarray}
  {\rm \bf CR2}                    & & \nonumber  \\
  {\rm   LO}  & : &  p^2,\ m_{\rm q}   \nonumber  \\
  {\rm  NLO}  & : &  a^2               \nonumber  \\
  {\rm NNLO}  & : &  p^4,\ p^2 m_{\rm q},\ m_{\rm q}^2,\ a^3\ .
\label{eqn:counting-rule2}
\end  {eqnarray}
and compare the results with CR2 
to those with CR1 given in the previous section. 
The CR2 corresponds to 
$m_{\rm q}\simeq a^{3/2}\Lambda_{\rm QCD}^{5/2} (=6.7)$ MeV
with $a^{-1}=2.19$ GeV and 
$\Lambda_{\rm QCD}=0.217(24)$ GeV 
in the $\overline{\rm MS}$ scheme~\cite{PDG}, 
while the CR1 does to 
$m_{\rm q}\simeq a\Lambda_{\rm QCD}^{2}(=22)$ MeV. 
The quark mass parameters 
corresponding to $m_\pi=0.17$, $0.30$, $0.41$, $0.57$, $0.71$ GeV 
gives $m_{\rm ud}^{\rm \overline{MS}}=3.5$, $12$, $24$, $46$, $67$ MeV, 
respectively~\cite{Aoki:2008sm}. 
For the data in $m_\pi\le 0.41$ GeV, which are used in our chiral analysis, 
it is not clear 
which counting rule is appropriate from these rough estimations. 
We need the quantitative comparison 
for the choice of the counting rule.

Due to the ${\cal O}(a^3)$ terms, 
the WChPT formulae given by 
Eqs.~(\ref{eqn:su3wchpt-p4-dcp1-cr1-a0pp2})--(\ref{eqn:su3wchpt-p4-dcp1-cr1-a0pk1})
are changed to 
\begin{eqnarray}
  a_0^{(2)} m_\pi
  & = &
  a_0^{(2)} m_\pi\, 
  |_{\, {}_{\rm  ChPT}}
  - 
  \left(
    c_2 + c_3\cdot\frac{a}{f_\pi^2}
  \right)\cdot
  \frac{a^2}{16\pi f_\pi^2}\ ,
\label{eqn:su3wchpt-p4-dcp1-cr2-a0pp2}
\\
  a_0^{(1)} m_K
  & = &
  a_0^{(1)} m_K\, 
  |_{\, {}_{\rm  ChPT}}
  - 
  \left(
    c_2+ c_3\cdot\frac{a}{f_K^2}
  \right)\cdot
  \frac{a^2}{16\pi f_K^2}\ ,
\label{eqn:su3wchpt-p4-dcp1-cr2-a0kk1}
\\
  a_0^{(3/2)} \mu_{\pi K}
  & = &
  a_0^{(3/2)} \mu_{\pi K}\, 
  |_{\, {}_{\rm  ChPT}}
  - 
  \left(
    c_2+ c_3\cdot\frac{a}{f_\pi f_K}
  \right)\cdot
  \frac{a^2}{8\pi f_\pi f_K}\cdot 
  \frac{\mu_{\pi K}^2}{m_\pi m_K}\ ,
\label{eqn:su3wchpt-p4-dcp1-cr2-a0pk3}
\\
  a_0^{(1/2)} \mu_{\pi K}
  & = &
  a_0^{(1/2)} \mu_{\pi K}\, 
  |_{\, {}_{\rm  ChPT}}
  - 
  \left(
    c_2+ c_3\cdot\frac{a}{f_\pi f_K}
  \right)\cdot
  \frac{a^2}{8\pi f_\pi f_K}\cdot 
  \frac{\mu_{\pi K}^2}{m_\pi m_K}\ ,
\label{eqn:su3wchpt-p4-dcp1-cr2-a0pk1}
\end  {eqnarray}
with an additional free parameter $c_3$. 
Here, 
$a_0^{(2  )} m_\pi     \, |_{\rm ChPT}$, 
$a_0^{(1  )} m_K       \, |_{\rm ChPT}$, 
$a_0^{(3/2)}\mu_{\pi K}\, |_{\rm ChPT}$ and 
$a_0^{(1/2)}\mu_{\pi K}\, |_{\rm ChPT}$ 
are the scattering length of the ChPT given by 
Eqs.(\ref{eqn:su3chpt-p4-dcp1-a0pp2}),
    (\ref{eqn:su3chpt-p4-dcp1-a0kk1}),
    (\ref{eqn:su3chpt-p4-dcp1-a0pk3}) and
    (\ref{eqn:su3chpt-p4-dcp1-a0pk1}), respectively.

\begin{figure}[htbp]
\begin{center}
\includegraphics[width=148mm]{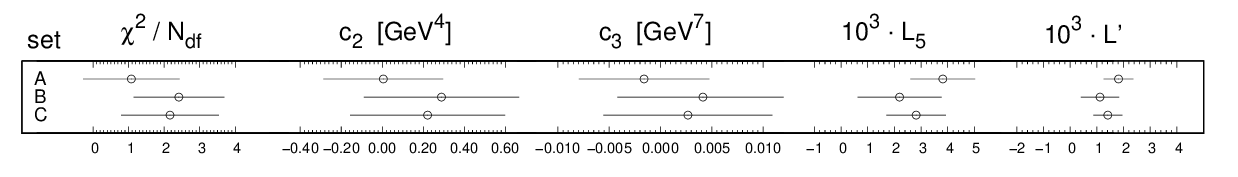}
\caption{
  $\chi^2/N_{\rm df}$ and
  LECs ($c_2$, 
        $c_3$, 
        $10^3\cdot  L_5     $ and 
        $10^3\cdot  L^\prime$)
  determined in the ${\cal O}(p^4)$ WChPT (CR2) fit 
  with the data sets A, B and C. 
}
\label{fig:su3wchpt-p4-cr2-summary}
\end  {center}
\end  {figure}

In Fig.\ref{fig:su3wchpt-p4-cr2-summary}, 
we show the results of the fitting.
$\chi^2/N_{\rm df}$ takes the reasonable value within the statistical error. 
The scattering lengths at the physical point 
obtained with the data set A, B and C are listed in 
Table.\ref{tbl:SL_at_physical_point_with_CR2}. 
For the data set B, they are given as 
\begin{eqnarray}
  a_0^{(2  )} m_\pi      &=& -0.04225(40)  \nonumber \\
  a_0^{(1  )} m_K        &=& -0.299  (31)  \nonumber \\
  a_0^{(3/2)}\mu_{\pi K} &=& -0.0461 (43)  \nonumber \\
  a_0^{(1/2)}\mu_{\pi K} &=&  0.145  (19)\ . 
\label{eqn:SCL_at_phys_point_with_CR2}
\end  {eqnarray}
These are consistent with those obtained with the CR1
in Eq.(\ref{eqn:SCL_at_phys_point}) 
and the systematic error caused by the choice of the counting rule
is negligible. 
Thus, 
we ignore the systematic error caused by the choice of the counting rule 
in the following discussion.

%
%----------------------
%
%----------------------
%

%
%----------------------
%
\begin{table}[p]
\begin{center}
\begin{tabular}{ccccc}
\hline
\hline
data set & \quad\quad & 
A              & B              & C               \\
\hline
$\chi^2/N_{\rm df}$ & & 
$ 1.1(1.4)$    & $ 2.4(1.3)$    & $ 2.2(1.4)$     \\

$c_2$ [GeV${}^4$] & & 
$ 0.01(29)$    & $ 0.29(38)$    & $ 0.22(38)$     \\
$c_3$ [GeV${}^7$] & & 
$-0.0016(64)$  & $ 0.0041(83)$  & $ 0.0027(82)$   \\
$10^3\cdot L_5$ & & 
$ 3.8(1.2)$    & $ 2.2(1.6)$    & $ 2.8(1.1)$     \\
$10^3\cdot L^\prime$ & & 
$ 1.81(56)$    & $ 1.12(72)$    & $ 1.41(54)$     \\
\hline
$a_0^{(2  )} m_\pi     $ & & 
$-0.04245(34)$ & $-0.04225(40)$ & $-0.04229(44)$  \\
\hline
$a_0^{(1  )} m_K       $ & & 
$-0.314  (26)$ & $-0.299  (31)$ & $-0.301  (33)$  \\
$a_0^{(1  )} m_\pi     $ & & 
$-0.0889 (72)$ & $-0.0846 (87)$ & $-0.0853 (94)$  \\
\hline
$a_0^{(3/2)}\mu_{\pi K}$ & & 
$-0.0504 (33)$ & $-0.0461 (43)$ & $-0.0476 (32)$  \\

$a_0^{(3/2)} m_\pi     $ & & 
$-0.0647 (43)$ & $-0,0592 (55)$ & $-0.0611 (41)$  \\
\hline
$a_0^{(1/2)}\mu_{\pi K}$ & & 
$ 0.164  (15)$ & $ 0.145  (19)$ & $ 0.153  (13)$  \\
$a_0^{(1/2)} m_\pi     $ & & 
$ 0.211  (19)$ & $ 0.186  (24)$ & $ 0.196  (17)$  \\
\hline
\hline
\end  {tabular}
\caption{
  $\chi^2/N_{\rm df}$ and LECs obtained 
  from the ${\cal O}(p^4)$ $SU(3)$ WChPT (CR2) fits 
  for the data sets A, B and C. 
  The scattering lengths at the physical point are also shown. 
}
\label{tbl:SL_at_physical_point_with_CR2}
\end  {center}
\end  {table}
%
%----------------------
%

%
% @@@ =======================================================================
%

\subsubsection{Finite volume}
\label  {sssec:finite_volume}

In this section, we discuss the systematic error of the finite volume, 
which appears from a deformation of the two-particle interaction 
due to the small lattice extent. 
For the $\pi\pi(I=2)$ system, 
the error has been estimated by the ${\cal O}(p^4)$ $SU(2)$ ChPT.
\cite{Bedaque:2006yi}. 
The contribution to $[\ \tan\delta_0(k)/k\ ]^{-1}$ 
is considered to be the order of ${\rm e}^{-m_\pi La}$. 
It is smaller than 
6\% of $[\ \tan\delta_0(k)/k\ ]^{-1}$ at $m_\pi=0.14$ GeV, 
and 1\% in $m_\pi\ge 0.29 $ GeV at $La=2.9$ fm. 
It is much smaller than our statistical errors. 
In the $SU(3)$ case, 
we need to consider the contributions due to the $K$ and $\eta$ meson. 
However, they are considered to be smaller than 
the contribution from the pion. 
Thus, we ignore this systematic error in the following discussion. 

%
% @@@ =======================================================================
%

\subsubsection{Uncertainty for $f_\pi$ and $f_K$}
\label  {sssec:Uncertainty_of_fpi_and_fkn}

We discuss the effects of the statistical 
uncertainty for the decay constants 
$f_\pi$ and $f_K$ in the WChPT formulae 
in Eqs.~(\ref{eqn:su3wchpt-p4-dcp1-cr1-a0pp2})--(\ref{eqn:su3wchpt-p4-dcp1-cr1-a0pk1}). 
In the following estimation, we use the data set B. 
In order to investigate the effects, 
we carry out the fitting with 
$(f_\pi\pm\sigma(f_\pi),f_K\pm\sigma(f_K))$ 
with the one standard errors, $\sigma(f_\pi)$ and 
%
%  "one standard deviations"  ->  "one standard errors"
%  This typo is modified in the version "fin2". 
%
$\sigma(f_K)$, 
whose values are tabulated in Table~\ref{tbl:fpi_fkn}. 
We regard the maximum absolute values of the differences 
among these fit results 
as the systematic error from the uncertainty of the decay constants. 
We obtain 
\begin{eqnarray}
  a_0^{(2  )} m_\pi      &=& -0.04243(22)\pm 0.00032  \nonumber \\
  a_0^{(1  )} m_K        &=& -0.312  (17)\pm 0.024    \nonumber \\
  a_0^{(3/2)}\mu_{\pi K} &=& -0.0477 (27)\pm 0.0020   \nonumber \\
  a_0^{(1/2)}\mu_{\pi K} &=&  0.150  (16)\pm 0.011\ , 
\end  {eqnarray}
where the second terms are the systematic errors from the decay constants. 
We find that these errors are comparable 
with the statistical error. 

%
% @@@ =======================================================================
%

\subsubsection{Higher order effects of ChPT}
\label  {sssec:higher_order_effects_of_ChPT}

In this section, 
we give rough estimations of contributions of the ${\cal O}(p^6)$ terms 
at the physical point. 
The ${\cal O}(p^2)$ and ${\cal O}(p^4)$ contributions of our results 
at the physical point are 
\begin{equation}
  \begin{array}{lllllll}
                           & & {\cal O}(p^2) & & {\cal O}(p^4) & \\
    a_0^{(2  )} m_\pi      &:& -0.04607      & & -0.04243(22)  & \\
    a_0^{(1  )} m_K        &:& -0.401        & & -0.312  (17)  & \\
    a_0^{(3/2)}\mu_{\pi K} &:& -0.0468       & & -0.0477 (27)  & \\
    a_0^{(1/2)}\mu_{\pi K} &:&  0.0936       & &  0.150  (16)  & .
  \end  {array}
\label{eqn:p6_effects_3}
\end  {equation}
We roughly estimate the pure ${\cal O}(p^6)$ contributions by 
$  X^{{}^{{\cal O}(p^2)}}\times 
(1-X^{{}^{{\cal O}(p^4)}}
  /X^{{}^{{\cal O}(p^2)}})^2$
for 
$X=a_0^{(2  )} m_\pi$, 
$  a_0^{(1  )} m_K  $, 
$  a_0^{(3/2)} \mu_{\pi K}$, 
$  a_0^{(1/2)} \mu_{\pi K}$. 
%
%  $  a_0^{(3/2)} m_\pi$, 
%  $  a_0^{(1/2)} m_\pi$. 
%
%  These typos are modified in the version "fin2". 
%
We regard them 
as the systematic error from an uncertainty 
of the higher order terms of ChPT. 
We obtain 
\begin{eqnarray}
  a_0^{(2  )} m_\pi      &=& -0.04243(22)\pm 0.00029  \nonumber \\
  a_0^{(1  )} m_K        &=& -0.312  (17)\pm 0.020    \nonumber \\
  a_0^{(3/2)}\mu_{\pi K} &=& -0.0477 (27)\pm 0.0001   \nonumber \\
  a_0^{(1/2)}\mu_{\pi K} &=&  0.150  (16)\pm 0.035\ , 
\end  {eqnarray}
where the second term refer to the systematic error.

This systematic error for the higher order effect 
is added to the systematic error due to the error of $f_\pi$ and $f_K$ 
in quadrature, and we regard it as the total systematic error 
which is given by the second term of Eq.(\ref{eqn:SCL_at_phys_point}).

%
% @@@ =======================================================================
%

\subsection{Comparison with the previous studies at the physical point}
\label{ssec:Comparison_with_the_previous_studies_at_the_physical_point}

For the $\pi K$ channels, 
some of the previous works used the values of 
$a_0^{(3/2)} m_\pi$ and 
$a_0^{(1/2)} m_\pi$, 
instead of 
$a_0^{(3/2)}\mu_{\pi K}$ and
$a_0^{(1/2)}\mu_{\pi K}$. 
For the comparison, we present these values of our results, 
\begin{eqnarray}
  a_0^{(3/2)} m_\pi &=& -0.0612(35)(26)  \nonumber \\
  a_0^{(1/2)} m_\pi &=&  0.193 (21)(47)\ . 
\end  {eqnarray}

\begin{figure}[htbp]
\begin{center}
\includegraphics[width=160mm]{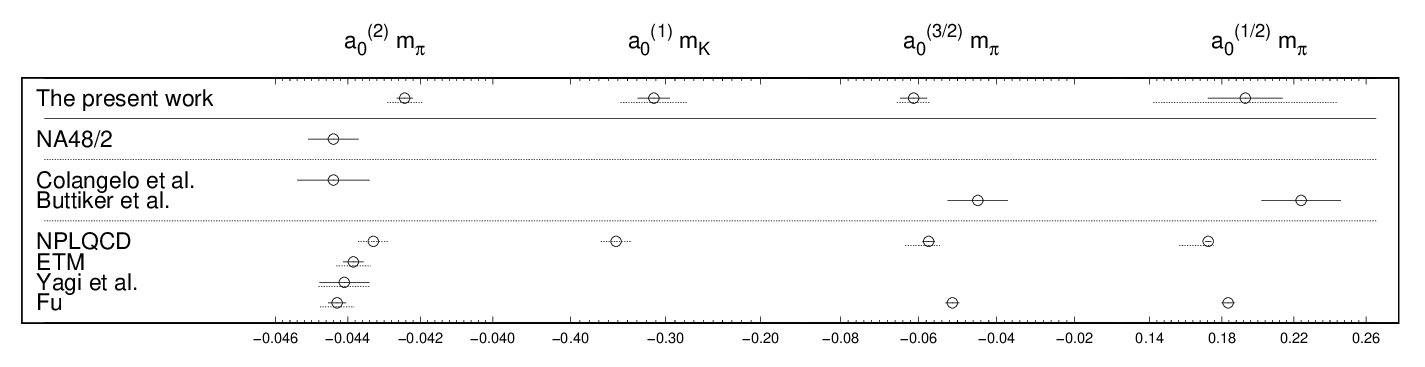}
\caption{
  $a_0^{(2  )} m_\pi$, 
  $a_0^{(1  )} m_K  $, 
  $a_0^{(3/2)} m_\pi$ and
  $a_0^{(1/2)} m_\pi$ in the present work are shown 
  in a comparison with the previous works. 
  As the previous works, 
  we refer 
  the experimental value by 
  NA48/2~\cite{Batley:2010zza}, 
  the phenomenological evaluations by 
  Colangelo  {\it et al.}~\cite{Colangelo:2000jc} and
  B\"uttiker {\it et al.}~\cite{Buettiker:2003pp}, and 
  the lattice calculations by 
  the NPLQCD Collaboration~\cite{NPLQCD.pp2,NPLQCD.KK1,NPLQCD.pK}, 
  the ETM    Collaboration~\cite{ETM.pp2}, 
  Yagi {\it et al.}~\cite{Yagi.pp2} and 
  Fu~\cite{Fu.pp.2,Fu.pK}. 
}
\label{fig:scl_su3wchpt_summary}
\end  {center}
\end  {figure}

In Table~\ref{tbl:SL_comparison} and Fig.~\ref{fig:scl_su3wchpt_summary}, 
we show 
$a_0^{(2  )} m_\pi$, 
$a_0^{(1  )} m_K  $, 
$a_0^{(3/2)} m_\pi$ and
$a_0^{(1/2)} m_\pi$ determined in the present work 
together with the previous works. 
As the previous works, 
we refer 
the experimental values by 
E865~\cite{Pislak:2003sv} and
NA48/2~\cite{Batley:2010zza}, 
the phenomenological evaluations by 
Colangelo  {\it et al.}~\cite{Colangelo:2000jc} and
B\"uttiker {\it et al.}~\cite{Buettiker:2003pp}, and 
the lattice calculations by 
the NPLQCD Collaboration~\cite{NPLQCD.pp2,NPLQCD.KK1,NPLQCD.pK}, 
the ETM    Collaboration~\cite{ETM.pp2},
Yagi {\it et al.}~\cite{Yagi.pp2} and 
Fu~\cite{Fu.pp.2,Fu.pK}. 
In the figure, we do not plot the result of
E865 due to the large statistical error. 
For the lattice calculations, 
we show the combined errors 
where the statistical and systematic errors are added in quadrature 
by dotted lines in addition to the statistical errors (solid lines).

Although all the lattice results in each channel are roughly consistent, 
there are the slight deviations from the previous works, 
especially in $a_0^{(2  )} m_\pi$. 
The reason for the deviations is not clear at the present. 
For the quantitative understanding, 
the systematic study with the different lattice spacings 
near the physical point is needed in the future.

%
%----------------------
%
%
%----------------------
%

%
%----------------------
%
\begin{table}[p]
\begin{center}
\begin{tabular}{ccccccccc}
\hline
\hline
                        & \quad\quad & 
$a_0^{(2  )} m_\pi$     & \quad\quad & 
$a_0^{(1  )} m_K  $     & \quad\quad & 
$a_0^{(3/2)} m_\pi$     & \quad\quad &
$a_0^{(1/2)} m_\pi$     \\
\hline
the present work        &  &
$-0.04243(22)(43)$      &  &
$-0.312  (17)(31)$      &  &
$-0.0612 (35)(26)$      &  &
$ 0.193  (21)(47)$      \\
\hline
E865~\cite{Pislak:2003sv} &  & 
$-0.0432  (86)$         &  & 
                        &  &
                        &  &
                        \\
NA48/2~\cite{Batley:2010zza} &  & 
$-0.0447  ( 7)$         &  & 
                        &  &
                        &  &
                        \\
\hline
Colangelo  {\it et al.}~\cite{Colangelo:2000jc} &  & 
$-0.0444  (10)$         &  & 
                        &  &
                        &  &
                        \\
B\"uttiker {\it et al.}~\cite{Buettiker:2003pp} &  & 
                        &  & 
                        &  &
$-0.0448(77)$           &  &
$ 0.224 (22)$           \\
\hline
NPLQCD~\cite{NPLQCD.pp2,NPLQCD.KK1,NPLQCD.pK} &  & 
$-0.04330(42)$          &  & 
$-0.352  (16)$          &  &
$-0.0574 (16)({}^{+24}_{- 58})$          &  &
$ 0.1725 (13)({}^{+23}_{-156})$          \\
ETM~\cite{ETM.pp2} &  &
$-0.04385 (28)(38)$     &  &
                        &  &
                        &  &
                        \\
Yagi {\it et al.}~\cite{Yagi.pp2} &  & 
$-0.04410 (69)(18)$     &  &
                        &  &
                        &  &
                        \\
Fu~\cite{Fu.pp.2,Fu.pK} &  & 
$-0.04430 (25)(40)$     &  &
                        &  &
$-0.0512  (18)$         &  &
$ 0.1819  (35)$         \\
\hline
\hline
\end  {tabular}
\caption{
  $a_0^{(2  )} m_\pi$, 
  $a_0^{(1  )} m_K  $, 
  $a_0^{(3/2)} m_\pi$ and
  $a_0^{(1/2)} m_\pi$ in the present work are shown 
  in a comparison with the previous works. 
  As the previous works, 
  we refer 
  the experimental value by 
  E865~\cite{Pislak:2003sv} and
  NA48/2~\cite{Batley:2010zza}, 
  the phenomenological evaluations by 
  Colangelo  {\it et al.}~\cite{Colangelo:2000jc} and
  B\"uttiker {\it et al.}~\cite{Buettiker:2003pp}, and 
  the lattice calculations by 
  the NPLQCD Collaboration~\cite{NPLQCD.pp2,NPLQCD.KK1,NPLQCD.pK}, 
  the ETM    Collaboration~\cite{ETM.pp2}, 
  Yagi {\it et al.}~\cite{Yagi.pp2} and 
  Fu~\cite{Fu.pp.2,Fu.pK}. 
  We note that 
  for $a_0^{(2)}m_\pi$ and $a_0^{(1)}m_K$ of the NPLQCD Collaboration, 
  the combined errors, 
  where the statistical and systematic errors are added in quadrature, 
  are listed. 
}
\label{tbl:SL_comparison}
\end  {center}
\end  {table}
%
%----------------------
%

%
% @@ ==================================================================
%
\section  {Conclusion}
\label{sec:Conclusion}

The interaction of the $S$-wave two-meson systems
($\pi\pi(I=2  )$, 
 $ K  K (I=1  )$, 
 $\pi K (I=3/2)$ and 
 $\pi K (I=1/2)$) has been studied from lattice QCD. 
To reduce the computational cost, we have employed the method 
where one of the particles in the final state is fixed at a given time. 
For the $\pi K(I=1/2)$ system, 
we have used the variational method with the two operators 
to separate the contamination from the higher states. 
We have observed that the interaction at low energy is 
repulsive  for the $\pi\pi(I=2  )$,
                   $ K  K (I=1  )$ and
                   $\pi K (I=3/2)$ systems, and 
attractive for the $\pi K (I=1/2)$ system. 
This feature is consistent with the experiment.

The scattering lengths have 
been calculated by using the L\"uscher's finite size method. 
We have found that the attraction in the $\pi K (I=1/2)$ system 
becomes so strong in $m_\pi> 0.41$ GeV 
that the sign of $\tan\delta_0(k)/k$ becomes negative. 
This fact indicates formation of a bound state 
at heavy $m_\pi$ for the $\pi K (I=1/2)$ system. 
Therefore, we have used the data in $m_\pi\le 0.30$ GeV 
to evaluate the reliable scattering length for this system.

We have investigated the quark mass dependence of the scattering lengths 
to evaluate the values at the physical quark mass. 
For this purpose, we have considered the ${\cal O}(p^4)$ ChPT formulae. 
However, the fitting with these formulae does not work 
for our results of the scattering length, especially at $m_\pi=0.17$ GeV. 
We alternatively have tried to fit 
with the ${\cal O}(p^4)$ WChPT formulae 
including the ${\cal O}(a^2)$ terms. 
We have found that these formulae reproduce 
the mass dependence of our results even near $m_\pi=0.17$ GeV. 
The description seems to work well at least 
in $m_\pi\le 0.41$ GeV for $a_0^{(2)} m_\pi$, 
in $m_\pi\le 0.30$ GeV for $a_0^{(1)} m_K$ and $a_0^{(3/2)} \mu_{\pi K}$, 
and 
in $m_\pi\le 0.30$ GeV for $a_0^{(1/2)} \mu_{\pi K}$. 
We have also discussed the possible systematic errors 
and evaluated the scattering lengths at the physical quark mass.

Although our lattice results are roughly consistent with 
the results of the previous studies, 
the deviations beyond the statistical error remain at the present. 
We need the systematic study with the different lattice spacings 
near the physical point for the quantitative understanding in the future. 

%
% @@@ =================================================================
%
\section*{Acknowledgments}
We thank colleagues in the PACS-CS Collaboration for helpful
discussions and providing us the code used in the present work. 
K. S. thanks to Y. Namekawa for reading the manuscript carefully. 
This study is supported by
Grants-in-Aid for Scientific Research on Priority Area (No. 21105506)
from the Ministry of Education, Culture, Sports, Science and Technology.
The numerical calculations were carried out 
on the super parallel computers, 
PACS-CS and T2K-Tsukuba at the University of Tsukuba, 
and TSUBAME at the Tokyo Institute of Technology. 
%
% @@ ==================================================================
%
\appendix

%
% @@ ==================================================================
%

\section  {${\cal O}(p^4)$ $SU(3)$ WChPT formulae}
\label{app:op4_su3_wchpt_formulae}

We give the formulae constructed from 
the ${\cal O}(p^4)$ $SU(3)$ WChPT with the CR1. 
According to Ref.\cite{Bar:2003mh}, 
the ${\cal O}(a^2)$ lagrangian consists of three terms written as 
\begin{equation}
  {\cal L}_{{\cal O}(a^2)}
  = w_6\cdot \frac{a^2 F^2}{16}\cdot \langle U  + U^\dagger    \rangle^2
  + w_7\cdot \frac{a^2 F^2}{16}\cdot \langle U  - U^\dagger    \rangle^2
  + w_8\cdot \frac{a^2 F^2}{8} \cdot \langle U^2+(U^\dagger)^2 \rangle 
\label{eqn:su3wchpt-1-lagrangian}
\end  {equation}
in the Minkowski space-time, 
where 
$U={\rm e}^{i\Phi/F}$ with the NG-boson field matrix $\Phi$, and 
the angle bracket means the trace for the flavor indices. 
$w_6$, $w_7$ and $w_8$ are the LECs in the $SU(3)$ WChPT. 
The LEC $c_2$ discussed in Sec. \ref{ssec:Chiral_analysis_with_op4_wchpt} 
is defined by $c_2\equiv - 8 w_6 - 4 w_8$. 
After calculating the generating function with zero external fields, 
we regard it as an effective action of NG boson fields 
according to Ref.~\cite{Gasser:1984pr}. 
On-shell quantities can be obtained from this effective action.

We represent 
the mass of the NG boson $P(=\pi,K,\eta)$ at the tree level by $M_P$. 
They are written as 
\begin{eqnarray}
  M_\pi^2   &=&  2B  m_{\rm ud}              \ ,  \nonumber  \\
  M_K^2     &=&   B( m_{\rm ud}+ m_{\rm s})  \ ,  \nonumber  \\
  M_\eta^2  &=&   B(2m_{\rm ud}+4m_{\rm s})/3\ , 
\end  {eqnarray}
with a parameter $B$, bare quark masses 
$m_{\rm ud}(\equiv m_{\rm u}=m_{\rm d})$ and $m_{\rm s}$. 
It is useful to consider the shifted mass with the ${\cal O}(a^2)$ terms as 
\begin{equation}
  \bar{M}_P^2=M_P^2+(12w_6+4w_8)a^2\ 
\end  {equation}
because the NG boson mass always enters 
the WChPT lagrangian with the form of $\bar{M}_P^2$.

The NG boson masses up to the ${\cal O}(p^4)$ terms 
can be described as 
\begin{eqnarray}
  m_\pi^2
  &=&
  \bar{M}_\pi^2
  \left[\ 
      1
    + \frac{\bar{M  }_\pi^2}{F^2}(- 8L_4-8L_5+16L_6+16L_8)
    + \frac{\bar{M  }_K^2  }{F^2}(-16L_4     +32L_6      )
    + \frac{\bar{\mu}_\pi  }{F^2}
    - \frac{1}{3}
      \frac{\bar{\mu}_\eta }{F^2}\ 
  \right]\ ,
  \nonumber  \\
  & & 
\label{eqn:su3wchpt_p4_mpi2_org}
\\
  m_K^2
  &=&
  \bar{M}_K^2
  \left[\ 
      1
    + \frac{\bar{M  }_K^2  }{F^2}(-16L_4-8L_5+32L_6+16L_8)
    + \frac{\bar{M  }_\pi^2}{F^2}(- 8L_4     +16L_6      )
    + \frac{2}{3}
      \frac{\bar{\mu}_\eta }{F^2}\ 
  \right]\ ,
\label{eqn:su3wchpt_p4_mkn2_org}
\end  {eqnarray}
where 
$\displaystyle
 \bar{\mu}_P
 =\frac{1}{32\pi^2}
 \bar{M  }_P^2 \log(\bar{M}_P^2/\mu^2)$\ .
The difference from the continuum ChPT originates only from 
$\bar{M}_P^2$ in the leading order. 
It is noted that $\bar{M_\pi}^4$, $\bar{M_\pi}^2\bar{M}_K^2$ and $\bar{M_K}^4$ 
in the next order are indistinguishable from 
$M_\pi^4$, $M_\pi^2 M_K^2$ and $M_K^4$, respectively, 
up to ${\cal O}(p^4)$ terms.

The decay constant of the pion and kaon up to the ${\cal O}(p^4)$ terms 
can be described as 
\begin{eqnarray}
  f_\pi
  &=&
  F
  \left[\ 
      1
    + \frac{\bar{M  }_\pi^2}{F^2}(4L_4+4L_5)
    + \frac{\bar{M  }_K^2  }{F^2}(8L_4)
    -2\frac{\bar{\mu}_\pi  }{F^2}
    - \frac{\bar{\mu}_K    }{F^2}\ 
  \right]\ ,
\label{eqn:su3wchpt_p4_fpi-F_org}
\\
  f_K
  &=&
  F
  \left[\
      1
    + \frac{\bar{M  }_K^2  }{F^2}(8L_4+4L_5)
    + \frac{\bar{M  }_\pi^2}{F^2}(4L_4)
    - \frac{3}{4}
      \frac{\bar{\mu}_\pi  }{F^2}
    - \frac{3}{2}
      \frac{\bar{\mu}_K    }{F^2}
    - \frac{3}{4}
      \frac{\bar{\mu}_\eta }{F^2}\ 
  \right]\ .
\label{eqn:su3wchpt_p4_fkn-F_org}
\end  {eqnarray}
The difference from the continuum ChPT does not exist 
because 
$\bar{M_\pi}^2$ and $\bar{M_K}^2$ in the next-leading order
are indistinguishable from 
$M_\pi^2$ and $M_K^2$, respectively, 
up to ${\cal O}(p^4)$ terms.

The scattering lengths of the 
$\pi\pi(I=2  )$, 
$ K  K (I=1  )$, 
$\pi K (I=3/2)$ and 
$\pi K (I=1/2)$ systems 
are already given in 
Eqs.(\ref{eqn:su3wchpt-p4-dcp1-cr1-a0pp2}), 
    (\ref{eqn:su3wchpt-p4-dcp1-cr1-a0kk1}), 
    (\ref{eqn:su3wchpt-p4-dcp1-cr1-a0pk3}) and 
    (\ref{eqn:su3wchpt-p4-dcp1-cr1-a0pk1}) 
with 
Eqs.(\ref{eqn:su3chpt-p4-dcp1-a0pp2}), 
    (\ref{eqn:su3chpt-p4-dcp1-a0kk1}), 
    (\ref{eqn:su3chpt-p4-dcp1-a0pk3}) and 
    (\ref{eqn:su3chpt-p4-dcp1-a0pk1}) 
except for the definitions of $\zeta^{(2),(1),(3/2),(1/2)}$. 
They are written as\footnote{
In the previous version of this article, 
the last term in the square bracket of Eq. (\ref{eqn:zeta1}) 
was given by $10 m_K^2/9$. 
However, the correct expression is $7 m_K^2/9$. 
We would like to thank A. Walker-Loud 
for pointing out the mistake in Eq. (\ref{eqn:zeta1}) 
and having a fruitful discussion \cite{Walker-Loud:2021}. 
Along with this correction, 
we have redone the chiral analysis. 
The LECs and the scattering lengths at the physical point 
are shifted at most by one standard errors, 
but we stress that the main argument is not changed. 
In the present version of this article, 
all results in the chiral analysis have been already revised. 
}
%  This footnote is added in the version "fin2". 
%
\begin{eqnarray}
  \zeta^{(2)}\hspace{3mm}
  &=&
  \frac{1}{(16\pi)^2}
  \left[\
  -\ \frac{ 3 m_\pi^2}{ 2}\log(\frac{m_\pi^2 }{\mu^2})\
  -\ \frac{   m_\pi^2}{18}\log(\frac{m_\eta^2}{\mu^2})\
  +\ \frac{ 4 m_\pi^2}{9}\
  \right]
\ ,
\\
  \zeta^{(1)}\hspace{3mm}
  &=&
  \frac{1}{(16\pi)^2}
  \left[\
     \frac{ m_\pi^2 m_K^2 }
          { 4 ( m_K^2 - m_\pi^2 )}
     \log(\frac{m_\pi^2}{\mu^2})
  \right.
  \nonumber  \\
  & &
  -\ m_K^2
     \log(\frac{m_K^2}{\mu^2})
  \nonumber  \\
  & &
  +\ \frac{ - 20 m_K^4 + 11 m_\pi^2 m_K^2 }
          { 36( m_K^2 - m_\pi^2 )}
     \log(\frac{m_\eta^2}{\mu^2})
  \nonumber  \\
  & &
  \left.
  +\ \frac{7 m_K^2}{9}\
  \right]
\ ,
\label{eqn:zeta1}
\\
  \zeta^{(3/2)}
  &=&
  \frac{1}{(16\pi)^2}
  \left[\
     \frac{ 22 m_\pi^3 m_K + 11 m_\pi^2 m_K^2 - 5 m_\pi^4}
          { 8 ( m_K^2 - m_\pi^2 )}
     \log(\frac{m_\pi^2}{\mu^2})
  \right.
  \nonumber  \\
  & &
  +\ \frac{ 9 m_K^4 - 134 m_\pi m_K^3 + 16 m_\pi^3 m_K - 55 m_\pi^2 m_K^2 }
          { 36( m_K^2 - m_\pi^2 )}
     \log(\frac{m_K^2}{\mu^2})
  \nonumber  \\
  & &
  +\ \frac{ 36 m_K^4 + 48 m_\pi m_K^3 - 10 m_\pi^3 m_K + 11 m_\pi^2 m_K^2 - 9 m_\pi^4 }
          { 72( m_K^2 - m_\pi^2 )}
     \log(\frac{m_\eta^2}{\mu^2})
  \nonumber  \\
  & &
  \left.
  +\ \frac{43 m_\pi m_K}{9}
  -\ \frac{ 8 m_\pi m_K}{9}\cdot t_1( m_\pi, m_K )
  \right]
\ ,
\\
  \zeta^{(1/2)}
  &=&
  \frac{1}{(16\pi)^2}
  \left[\
  +\ \frac{ 11 m_\pi^3 m_K - 11 m_\pi^2 m_K^2 + 5 m_\pi^4}
          { 4 ( m_K^2 - m_\pi^2 )}
     \log(\frac{m_\pi^2}{\mu^2})
  \right.
  \\
  & &
  +\ \frac{ - 9 m_K^4 - 67 m_\pi m_K^3 + 8 m_\pi^3 m_K + 55 m_\pi^2 m_K^2 }
          { 18( m_K^2 - m_\pi^2 )}
     \log(\frac{m_K^2}{\mu^2})
  \nonumber  \\
  & &
  +\ \frac{ - 36 m_K^4 + 24 m_K^3 m_\pi - 5 m_K m_\pi^3
            - 11 m_K^2 m_\pi^2 + 9 m_\pi^4 }
          { 36( m_K^2 - m_\pi^2 )}
     \log(\frac{m_\eta^2}{\mu^2})
  \nonumber  \\
  & &
  \left.
  +\ \frac{43 m_\pi m_K}{9}
  +\ \frac{ 4 m_\pi m_K}{9}\cdot t_1( m_\pi, m_K )
  -\ \frac{12 m_\pi m_K}{9}\cdot t_2( m_\pi, m_K )
  \right]
\ ,
\end  {eqnarray}
where $t_1( m_\pi, m_K )$, $t_2( m_\pi, m_K )$ can be written as 
\begin{eqnarray}
  t_1( m_\pi, m_K )
  &=&
  \frac{ \sqrt{( m_K + m_\pi )( 2 m_K - m_\pi )} }{ m_K - m_\pi }
  \arctan
  \left(
    \frac{ 2( m_K - m_\pi )}{ m_K + 2 m_\pi }
    \sqrt{
    \frac{ m_K + m_\pi }{ 2 m_K - m_\pi }
    }
  \right)
\ ,
\\
  t_2( m_\pi, m_K )
  &=&
 \frac{ \sqrt{( m_K - m_\pi )( 2 m_K + m_\pi )} }{ m_K + m_\pi }
  \arctan
  \left(
    \frac{ 2( m_K + m_\pi )}{ m_K - 2 m_\pi }
    \sqrt{
    \frac{ m_K - m_\pi }{ 2 m_K + m_\pi }
    }
  \right)
\ .
\end  {eqnarray}
We used $\displaystyle\bar{M  }_P^2=  m_P^2$, 
$\displaystyle\bar{\mu}_P^2=\mu_P^2
 \left(\equiv \frac{1}{32\pi^2}m_P^2\log(m_P^2/\mu^2)\right)$ and 
$F^2=f_\pi^2=f_K^2$ at the ${\cal O}(p^4)$ terms 
to simplify the expression. 
We also used the tree-level (Gell-Mann-Okubo) relation, 
$m_\eta^2=(4m_K^2-m_\pi^2)/3$ for the mass of the $\eta$ meson. 
These relations are sufficient 
if we restrict ourselves up to the ${\cal O}(p^4)$ expression. 

%
% @@ ==================================================================
%

%
\end{document}